\documentclass[aps,pra,twocolumn,nofootinbib,floatfix,superscriptaddress]{revtex4-2}
\usepackage{amsmath,mleftright,mathtools}
\usepackage{bm}
\usepackage{stix,microtype}
\usepackage{graphicx}
\usepackage{xspace}
\usepackage{units}
\usepackage[colorlinks,linkcolor=blue,citecolor=blue,urlcolor=blue]{hyperref}
\usepackage[dvipsnames]{xcolor}
\usepackage{cleveref}
\usepackage{array}   % for \newcolumntype macro
\newcolumntype{L}{>{$}c<{$}} % math-mode version of "l" column type
\newcolumntype{C}{>{$}c<{$}} % math-mode version of "l" column type
\usepackage{dcolumn}
\newcommand{\Hh}{\hat{H}}

\newcommand{\hphi}{\hat{\phi}}

\newcommand{\hd}{\hat{d}}
\newcommand{\ha}{\hat{a}}
\newcommand{\hx}{\hat{x}}
\newcommand{\hp}{\hat{p}}
\newcommand{\hA}{\hat{A}}
\newcommand{\s}{\sigma}
\graphicspath{{./Figs/}}
\mathchardef\mhyphen="2D

\newcommand{\be}{\begin{equation}}
\newcommand{\ee}{\end{equation}}

\newcommand{\Tr}{\text{Tr}}

\newcommand{\mv}{\mathbfit{v}}

\newcommand{\mh}{\mathbfit{h}}

\newcommand{\ma}{\bm{\alpha}}
\newcommand{\mO}{\bm{\Omega}}
\newcommand{\mD}{\mathbfit{D}}

\newcommand{\mPsi}{\bm{\Psi}}
\newcommand{\mGamma}{\bm{\Gamma}}

\newcommand{\mSgm}{\mathbfit{\Sigma}}

\newcommand{\mphi}{\bm{\phi}}

\newcommand{\cG}{\mathcal{G}}

\newcommand{\mcG}{\bm{\mathcal{G}}}
\newcommand{\mI}{\mathbfit{I}}

\newcommand{\mrho}{\bm{\rho}}

\newcommand{\bgg}{\bm{\gamma}}

\newcommand{\mgamma}{\bm{\gamma}}

\def\bgm{\mbox{\boldmath $\mu$}}

\def\bga{\mbox{\boldmath $\alpha$}}

\def\bgg{\mbox{\boldmath $\gamma$}}
\def\bcallg{\mbox{\boldmath $g$}}

\keywords{Nonequilibrium Green's function theory, generalized Kadanoff-Baym Ansatz, excited states}
\begin{document}
\title{Time-linear scaling NEGF method for real-time simulations of 
interacting electrons and bosons. II. Dynamics of polarons and doublons}
\author{Y. Pavlyukh}
\affiliation{Dipartimento di Fisica, Universit{\`a} di Roma Tor Vergata, Via della Ricerca Scientifica 1,
00133 Rome, Italy}
%\email{yaroslav.pavlyukh@gmail.com}
\author{E. Perfetto}
\affiliation{Dipartimento di Fisica, Universit{\`a} di Roma Tor Vergata, Via della Ricerca Scientifica 1,
00133 Rome, Italy}
\affiliation{INFN, Sezione di Roma Tor Vergata, Via della Ricerca Scientifica 1, 00133 Rome, Italy}
\author{Daniel Karlsson}
\affiliation{Department of Physics, Nanoscience Center P.O.Box 35
FI-40014 University of Jyv\"{a}skyl\"{a}, Finland}
\author{Robert van Leeuwen}
\affiliation{Department of Physics, Nanoscience Center P.O.Box 35
FI-40014 University of Jyv\"{a}skyl\"{a}, Finland}
\author{G. Stefanucci}
\affiliation{Dipartimento di Fisica, Universit{\`a} di Roma Tor Vergata, Via della Ricerca Scientifica 1,
00133 Rome, Italy}
\affiliation{INFN, Sezione di Roma Tor Vergata, Via della Ricerca Scientifica 1, 00133 Rome, Italy}
\date{\today}
\begin{abstract}
Nonequilibrium dynamics of the open chain Holstein-Hubbard model is studied using the
linear time-scaling GKBA+ODE scheme developed in the preceeding paper. We focus on the set
of parameters relevant for photovoltaic materials, i.\,e., a pair of electrons interacting
with phonons at the cross-over between the adiabatic and anti-adiabatic regimes and at
moderately large electron-electron interaction.  By comparing with exact solutions for two
corner cases, we demonstrate the accuracy of the $T$-matrix (in the $pp$ channel) and the
second-order Fan ($GD$) approximations for the treatment of electronic ($e$-$e$) and
electron-phonon ($e$-ph) correlations, respectively. The feedback of electron on phonons
is consistently included and is shown to be mandatory for the total energy
conservation. When two interactions are simultaneously present, our simulations offer a
glimpse into the dynamics of doublons and polarons unveiling the formation, propagation
and decay of these quasiparticles, energy redistribution between them and self-trapping of
electrons.
\end{abstract}
\maketitle
%===== ===== ===== ===== ===== =====   I   ===== ===== ===== ===== ===== =====
\section{Introduction}                            
%===== ===== ===== ===== ===== ===== ===== ===== ===== ===== ===== ===== =====
Even in the absence of $e$-$e$ interactions, electrons coupled to bosons represent one of
the most studied systems in
physics~\cite{mahan_many-particle_2000,van_leeuwen_first-principles_2004,
  giustino_electron-phonon_2017}. Restricting to solid state and molecular physics, there
are numerous models of increasing complexity: a core electron coupled to
plasmons~\cite{langreth_singularities_1970, schuler_time-dependent_2016,
  pavlyukh_pade_2017}, the Holstein
dimer~\cite{sakkinen_many-body_2015,tuovinen_phononic_2016} or a quantum dot coupled to
leads~\cite{galperin_non-linear_2008,leturcq_franckcondon_2009,white_inelastic_2012,
  perfetto_image_2013,wilner_nonequilibrium_2014,
  muhlbacher_real-time_2008,galperin_photonics_2017} as a paradigmatic model for the
Franck-Condon blockade in nanoscale transport, and infinite 1D and 2D systems at various
fillings representing phenomena ranging from the energy transfer along conjugated polymer
chains and photovoltaic devices~\cite{chen_dynamics_2015, tanimura_numerically_2020}, the
oscillations in the excitonic condensate in transition metal di- and
ternary-chalcogenides~\cite{werdehausen_coherent_2018}, to correlated cuprate
systems~\cite{kemper_direct_2015, stolpp_charge-density-wave_2020}. A common physical
phenomenon pertinent to all of them is the emergence of a new kind of
quasiparticle\,---\,the polaron\,---\,an electron surrounded by a cloud of coherent
phonons~\cite{devreese_frohlich_2009}.

Historically, one-dimensional coordination polymers~\cite{leong_one-dimensional_2011} were
among the first systems where $e$-ph dynamics have been studied
experimentally~\cite{dexheimer_femtosecond_2000, sugita_wave_2001} and
theoretically~\cite{ku_quantum_2007}, and a good understanding of polaron formation and
localization has been obtained.  In fact, they are convenient paradigmatic materials for
theoretical NEGF investigations. They can be characterized by a small set of parameters,
and the 1D nature makes them amenable to alternative theoretical methods, including the
wave-function~\cite{ku_quantum_2007,golez_relaxation_2012,golez_dissociation_2012,dorfner_real-time_2015},
the density-matrix renormalization group~\cite{rausch_filling-dependent_2017}, the
hierarchical equations of motion~\cite{chen_dynamics_2015} and matrix product
state~\cite{kloss_multiset_2019,frahm_ultrafast_2019} based ones.  Much more versatile and
practically relevant are materials for photovoltaic
applications~\cite{ponseca_ultrafast_2017}. By the very nature of solar energy conversion,
several crucial aspects\,---\,creation of nonequilibrium carriers, their dressing and
formation of the polaronic quasiparticles, and transport through the active material to
electrodes\,---\,creates a good predisposition for a NEGF theory and poses very
interesting challenges, e.\,g., simulation of picosecond (ps) polaron self-localization in
$\alpha$-Fe$_2$O$_3$ photoelectrochemical cells~\cite{pastor_situ_2019}.

We are close to achieving the goal of ps dynamics within a NEGF theory: $e$-ph
thermalization has already been demonstrated in a model insulator without $e$-$e$
interactions on a 2\,ps time-scale~\cite{karlsson_fast_2021}, whereas $e$-$e$ interacting
2D systems have been propagated for 100\,fs~\cite{perfetto_tdgw_2021}. Here, we illustrate
our methods by applying them to electron-phonon ($e$-ph) dynamics in the 1D
Holstein-Hubbard model~\cite{werner_phonon-enhanced_2013,kloss_multiset_2019}. The model
features important physical mechanisms present in realistic systems and enables us to
benchmark the GKBA+ODE scheme of the real-time NEGF theory in complicated cross-over
regimes.

The outline of our work is as follows.  After recapitulating basic ingredients of the NEGF
formalism (see paper I for the full-fledged theory) in Section~\ref{sec:derivation}, we
introduce the open-chain Holstein-Hubbard model in Section~\ref{sec:numsim}. Numerical
scaling with the system dimension and strategies for code optimization are discussed and
compared with alternative approaches.  More insights into the model dynamics are obtained
in the partial cases where only $e$-$e$ or $e$-ph interaction is present. Therefore, in
Sec.~\ref{sec:e-e} we provide benchmarks against numerically exact solutions of the
Hubbard model. Besides finding a high level of accuracy of the $T$-matrix approximation in
the particle-particle channel, it is shown that doublons play an important role in the
system dynamics.  In Sec.~\ref{sec:e-ph} the concept of polaron is introduced and results
for the electron localization in the Holstein model are presented. Then, we discuss how
electrons and doublons propagate when $e$-$e$ and $e$-ph are simultaneously present
(Sec.~\ref{sec:both}) at the level of $T$-matrix and the fully dressed second-order Fan
($GD$) approximation.  We find that $e$-ph interactions modify the localization of
doublons in a nontrivial way. We discuss two competing mechanisms and support our
interpretation by analyzing total energy contributions and $e$-ph correlators. Conclusions
and outlook are drawn in Section~\ref{sec:summary}.

%===== ===== ===== ===== ===== =====   II  ===== ===== ===== ===== ===== =====
\section{Electron-boson NEGF equations}%
\label{sec:derivation}                               
%===== ===== ===== ===== ===== ===== ===== ===== ===== ===== ===== ===== =====
The theory developed in the preceding paper I is completely general and can be applied to
a wide range of physical systems out of equilibrium with electron-electron and
electron-boson interactions. Here we consider specifically electrons interacting with
phonons as one of the most interesting and technologically relevant cases. Henceforth we
use latin letters to denote one-electron states; thus $i\equiv(\bm{i},\sigma)$ is a
composite index standing for an orbital degree of freedom $\bm{i}$ and a spin projection
$\s$. For phonons, one may work with standard creation and annihilation operators
$\ha_{\bm{\mu}}^\dagger$ and $\ha_{\bm{\mu}}$ for the mode $\bm{\mu}$,
respectively. However, since electrons typically couple to the phononic displacement, it
is of advantage to introduce operators of displacement and momentum
\begin{align}
  \hx_{\bm{\mu}}&=\frac{1}{\sqrt{2}}\left(\ha_{\bm{\mu}}^\dagger+\ha_{\bm{\mu}}\right),&
  \hp_{\bm{\mu}}&=\frac{i}{\sqrt{2}}\left(\ha_{\bm{\mu}}^\dagger-\ha_{\bm{\mu}}\right).
  \label{eq:def:x:p}
\end{align}
The composite greek index $\mu=(\bgm,\xi)$ specifies the phononic mode and the component
of the vector.  It is convenient to introduce two-components operators, with components
distinguished by a pseudospin $\xi$
\begin{align}
  \hA_\mu&=\begin{pmatrix}
    \ha_{\bm{\mu}}^\dagger\\\ha_{\bm{\mu}}
  \end{pmatrix}_\xi,&
  \hphi_\mu&=\begin{pmatrix}
    \hx_{\bm{\mu}}\\\hp_{\bm{\mu}}
  \end{pmatrix}_\xi.
\end{align}
From standard commutation rules,
\begin{align}
  \left[\ha_{\bm{\mu}},\ha_{\bm{\mu}'}^\dagger\right]&=\delta_{\bm{\mu} \bm{\mu}'},&
  \left[\hx_{\bm{\mu}},\hp_{\bm{\mu}'}\right]&=i\delta_{\bm{\mu} \bm{\mu}'},
\end{align}
it is not difficult to derive the commutation relations for two-component operators:
\begin{align}
  \left[\hA_\mu^\dagger,\hA_{\mu'}\right]&=\delta_{\bm{\mu} \bm{\mu}'}\sigma^{(3)}_{\xi\xi'},\;
  \left[\hphi_\mu,\hphi_{\mu'}\right]=-\delta_{\bm{\mu} \bm{\mu}'}\sigma^{(2)}_{\xi\xi'}
  =\alpha_{\mu\mu'}. 
\end{align}
Here, $\sigma^{(i)}$, $i=1,\ldots3$ are the pseudospin Pauli matrices. Therefore, the
noninteracting phononic Hamiltonian can be written as a quadratic form:
\begin{subequations}
  \label{eq:Hph}
\begin{align}
  \Hh_\text{ph}&=\sum_{\bm{\mu}}\omega_{\bm{\mu}}\left(\ha_{\bm{\mu}}^\dagger\ha_{\bm{\mu}}+\frac12\right)=
  \hA^\dagger \bm{\omega} \hA=\hphi^\dagger\bm{\Omega}\hphi,\\
  \bm{\omega}&=\bm{\Omega}=\mathrm{diag}(\omega)\otimes\frac12
  \begin{pmatrix}
    1&0\\0&1
  \end{pmatrix}.
\end{align}
\end{subequations}

The electron-phonon interaction is written in the form
% ---  
\begin{align}\label{eq:H:e:b}
 \Hh_\text{$e$-ph}(t)&= \sum_{\mu, ij} g_{\mu, ij}(t)\hd_{i}^\dagger \hd_{j}\hphi_\mu.
\end{align}
Density functional perturbation theory~\cite{baroni_phonons_2001,marini_many-body_2015} is
a well-established tool to construct the $e$-ph coupling matrix
elements~\cite{giustino_electron-phonon_2017} ($g_{\mu,ij}$ tensor) from first principles.

In the NEGF formalism the fundamental unknowns are the electronic lesser/greater
single-particle Green's functions $G^{\lessgtr}$ and their phononic counterparts
$D^{\lessgtr}$.  They satisfy the integro-differential Kadanoff-Baym equations (KBE) of
motion. The KBE can also be used to generate the EOMs for the electronic
$\rho_{ij}^<=\langle \hd_{j}^\dagger \hd_{i}\rangle$ and phononic
$\gamma^<_{\mu\nu}=\langle \hphi_\nu \hphi_\mu\rangle - \langle \hphi_\nu\rangle\langle
\hphi_\mu\rangle$ density matrices
\begin{align}
  \frac{d}{dt}\rho^<(t)&=-i\big[h^{e}(t),\rho^<(t) \big] 
  -\left(I^{e}(t)+\left(I^{e}(t)\right)^\dagger\right), \label{eq:eomrho:e}\\
  \frac{d}{dt}\bgg^<(t)&=-i \big[\mh^\text{ph}(t),\bgg^<(t) \big]
  +\left(\mI^\text{ph}(t)+\left(\mI^{\text{ph}}(t)\right)^\dagger\right).
  \label{eq:eomrho:b}
\end{align}
The effective electron Hamiltonian $h^e$ is a sum of the mean-field electronic
$h_{\text{HF}}(t)=h(t)+V_{\text{HF}}(t)$ and phononic
$h^{\text{$e$-ph}}(t)=\sum_{\mu}g_{\mu}\langle \hat\phi_\mu(t)\rangle$ parts, i.\,e.,
$h^e(t)=h_{\text{HF}}(t)+h^{\text{$e$-ph}}(t)$. $\mh^\text{ph}\equiv \bga(\mO+\mO^T)$ is
the effective phononic Hamiltonian.

\begin{figure}[tbp]
\centering  \includegraphics[width=0.82\columnwidth]{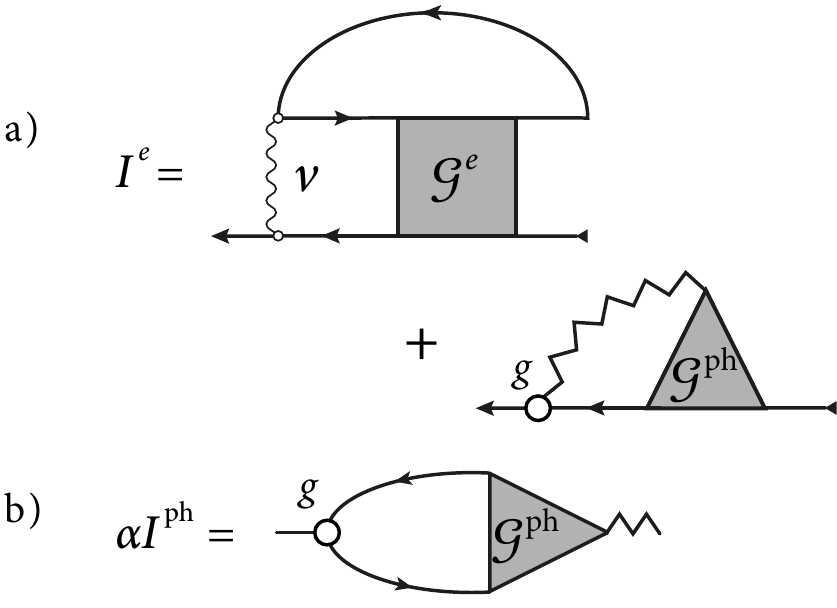}
\caption[]{Diagrammatic representation of the collision integrals in terms of high order
  Green's functions. Full lines are used for $G$ and zig-zag lines are used for
  $D$.\label{fig:collisions}}
\end{figure}

Equations~\eqref{eq:eomrho:e} and \eqref{eq:eomrho:b} are not closed because the collision
integrals $I^{e}$ and $\mI^\text{ph}$ are proportional to the more complicated correlators
$\mcG^{e}$ and $\mcG^\text{ph}$, respectively, see Fig.~\ref{fig:collisions}. Explicitly
\begin{align}
  \cG^{e}_{imjn}&=-\langle\hd_{n}^\dagger\hd_{j}^\dagger \hd_{i} \hd_{m}\rangle_c,&
  \cG^\text{ph}_{\mu,ij}&=\langle\hd_{j}^\dagger \hd_{i}\hphi_\mu \rangle_c.
\end{align}  
In paper~I we have shown using the generalized Kadanoff-Baym ansatz
(GKBA)~\cite{lipavsky_generalized_1986} that they, in turn, fulfill their own equations of
motions: for electrons in 2B and $T^{pp}$
approximations~\cite{schlunzen_achieving_2020,pavlyukh_photoinduced_2021}
\begin{multline}
  i \frac{d}{dt}\mcG^{e}(t)=-\mPsi^{e}(t)+\left[\mh^{e}(t) 
   +a\mrho^\Delta(t)\mv(t)\right]\mcG^{e}(t)\\
  -\mcG^{e}(t) \left[\mh^{e}(t) +a\mv(t) 
    \mrho^\Delta(t)\right].
  \label{eq:G2:X:init}
\end{multline}
And, similarly, for phonons~\cite{karlsson_fast_2021}
\begin{align}
  i\frac{d}{dt}\mcG^\text{ph}(t)&=-\mPsi^\text{ph}(t)+\mh^\text{ph}(t)\mcG^\text{ph}(t)-\mcG^\text{ph}(t)\mh^{e}(t),
\label{eq:EOM:gb}
\end{align}
We consistently use bold symbols to denote rank-2 matrices with phononic indices
(viz. Eq.~\ref{eq:eomrho:b} and \ref{eq:EOM:gb}) or electronic superindices
(Eq.~\ref{eq:G2:X:init}).  The constant $a$ in Eq.~\eqref{eq:G2:X:init} is equal to 0 for
the second Born (2B) approximation and to 1 for the $T$-matrix approximation in the $pp$
channel ($T^{pp}$).  In these equations $\mv$ is the Coulomb tensor,
$\mrho^\Delta=\mrho^>-\mrho^<$, and
\begin{subequations}
\begin{align}
  \mPsi^{e}(t)&\equiv \mrho^{>}(t)\mv(t)\mrho^{<}(t)-\mrho^{<}(t)\mv(t) \mrho^{>}(t),
  \label{psie}\\
  \mPsi^\text{ph}(t)&\equiv \bgg^{>}(t)\bcallg(t)\mrho^{<}(t)-\bgg^{<}(t)\bcallg(t) \mrho^{>}(t),
  \label{Psi:b:def}
\end{align}
\end{subequations}
are the driving terms.  We refer to paper I for the details related to the index order of
these quantities. To close the KBE one additionally needs to propagate the position and
momentum expectation values $\phi_\nu(t)=\langle \hat\phi_\nu(t)\rangle$ because they
enter $h^e$
\begin{align}\label{eq:phi:eom}
  i \frac{d}{dt} \phi_\mu(t) 
  -\sum\nolimits_{\nu}h_{\mu\nu}^\text{ph}(t)\phi_\nu(t) 
  = \sum\nolimits_{\nu ij}\alpha_{\mu\nu}g_{\nu,ij} \rho_{ji}(t).
\end{align}

Equations (\ref{eq:EOM:gb}) and (\ref{eq:G2:X:init}) together with EOMs for the electronic
and phononic density matrices [Eqs.~(\ref{eq:eomrho:e}) and (\ref{eq:eomrho:b})] and the
equation of motion for $\phi(t)$ [Eq.~(\ref{eq:phi:eom})], form a closed system of
ordinary differential equations (ODE).
%===== ===== ===== ===== ===== =====   IV  ===== ===== ===== ===== ===== =====
\section{1D Holstein-Hubbard model out of equilibrium}
\label{sec:numsim}
%===== ===== ===== ===== ===== ===== ===== ===== ===== ===== ===== ===== =====
The most demanding part of the GKBA+ODE calculations is the solution of the electronic
two-particle equation~\eqref{eq:G2:X:init}, whereby the matrix product $\mv \mcG$ scales
as $\mathcal{O}(N_\text{$e$-basis}^6)$, with the number of electronic basis functions
$N_\text{$e$-basis}$. The scaling can be improved relatively easy by imposing additional
physically motivated restrictions on the matrix elements of $\mv$. This brings us to the
consideration of the \emph{extended Hubbard model}.

In the extended Hubbard model, which is the target of our numerical implementation, there
are two classes of Coulomb integrals expressed in a site basis
\begin{align}
  v^d_{ij}&=v_{ijji},&
  v^x_{ij}&=v_{ijij}.\label{eq:assume}
\end{align}
All other Coulomb matrix elements are set to zero. This reduces the complexity of the
correlated GKBA methods to $\mathcal{O}(N_\text{$e$-basis}^5)$. In our implementation we
exploit this simplification whenever condition~\eqref{eq:assume} is fulfilled. The
implementation is otherwise completely general and can be used, for instance, to study
screening at the level of the $GW$
approximation~\cite{perfetto_tdgw_2021,bittner_photoinduced_2021,golez_multiband_2019,pavlyukh_photoinduced_2021}.

In this work, we focus on the dynamics of electrons and phonons in the 1D one-band
Holstein-Hubbard (h-h) model, where each site is coupled to a phonon,
\begin{multline}
  \hat{H}_\text{h-h}=-h\sum_{\sigma=\uparrow,\downarrow}\sum_{<\bm{i},\bm{j}>}\hd_{\bm{i}\sigma}^\dagger \hd_{\bm{j}\sigma}
  +U\sum_{\bm{i}}\hat{n}_{\bm{i}\uparrow}\hat{n}_{\bm{i}\downarrow}\\
  +\sum_{\bm{i}}\left\{\omega \ha^\dagger_{\bm{i}}\ha_{\bm{i}}+g(\ha^\dagger_{\bm{i}}+\ha_{\bm{i}})\hat{n}_{\bm{i}}\right\}.
  \label{eq:h:h-h}
\end{multline}
Here $h$ is the matrix element of the nearest neighbor hopping ($<\bm{i},\bm{j}>$ are the
neighboring sites), $U$ is the onsite Hubbard repulsion (therefore $v^d_{\bm{ij}} =
U\delta_{\bm{ij}}$ and $v^x=0$ in Eq.~\ref{eq:assume}), $\sigma$ is the spin projection,
$\omega$ is the frequency of phonons equal at all sites, i.\,e. $\omega_{\bm{\mu}}=\omega$
in Eqs.~\eqref{eq:Hph}, and $g$ is the coupling matrix element of the phonon displacement
at site $\bm{i}$ to the total electron density at the same site
($\hat{n}_{\bm{i}}=\sum_\sigma \hat{n}_{\bm{i}\sigma}$,
$\hat{n}_{\bm{i}\sigma}=\hd_{\bm{i}\sigma}^\dagger \hd_{\bm{i}\sigma}$), explicitly
$g_{\mu,ij}=\sqrt{2}g\delta_{\bm{i}\bm{j}}\delta_{\bm{\mu}\bm{i}}\delta_{\xi,1}$ in
Eq.~\eqref{eq:H:e:b}, where the $\sqrt{2}$ prefactor is in accordance with the definition
of the phononic displacement in Eq.~\eqref{eq:def:x:p}. The dynamics is triggered by the
creation of an electron (or a pair of them with zero total spin) at a given lattice
site. The lattice is otherwise empty. Here, we denote
$N=N_\text{$e$-basis}=N_\text{$b$-basis}$\,---\,the number of single particle electronic
and phononic basis functions. In what follows we measure energies in the units of $h$ and
times in the units of $1/h$, therefore the system is characterized by a set of three
parameters $\{U,\,\omega,\,g\}$. For the ease of reporting total energies, we eliminated
the zero-point vibrational energy $\tfrac12N\omega$ from the Hamiltonian~\eqref{eq:h:h-h}
in comparison with Eq.~\eqref{eq:Hph}. 

\begin{table}[]
  \caption{\label{tab:2} Largest systems and computational resources (Xeon Gold 5218 CPU @
    2.30GHz) used in this work.}
  \begin{ruledtabular}
    \begin{tabular}{cccrcdd}
      System & \multicolumn{2}{c}{Correlations} & State & Time & \multicolumn{2}{c}{CPU hours}\\\cline{2-3}\cline{6-7}
      $N$& $e$-$e$  & $e$-ph& vector & $t_\text{f}$ &\mbox{$e$-$e$}  &\mbox{$e$-ph}\\\hline \renewcommand{\arraystretch}{1.4}
      
      151         & HF & Ehrenfest &   23\,254   &  40 &0.02   & 0.07\\
      151         & HF & $GD$      & 7\,046\,264   &  40 &3.8    &  2.0\\
      151         & 2B & $GD$      & 526\,954\,666 &  40 &181.2  &  2.0\\
      151         & $T^{pp}$ & Ehrenfest & 519\,931\,656 &40 & 482.8 & 0.04\\
      151         & $T^{pp}$ & $GD$ & 526\,954\,666 &  40 &488.0 &   2.6\\
       \end{tabular}
    \end{ruledtabular}
\end{table}

The scenario with a \emph{single electron} has been studied in a number of works using
different methods.  Fehske \emph{et al.}~\cite{fehske_spatiotemporal_2011} used the
Chebyshev expansion technique to solve the Schr\"{o}dinger equation in a truncated bosonic
basis and to study the polaronic cloud formation characterized by the diagonal $\langle
\hat{n}_{\bm{i}}(\ha^\dagger_{\bm{i}+\bm{x}}+\ha_{\bm{i}+\bm{x}})\rangle$ ($\bm{x}=0$)
correlator, and the phonon emission and reabsorption processes leading to the electron
effective mass enhancement in a smaller system comprising only 17
sites. Interference/reflection of the electron wave-packet from the boundaries prevent
clear interpretation at larger times. Similar a wave-function (WF) approach has been used
by Gole\v{z} \emph{et al.}~\cite{golez_relaxation_2012} to study the relaxation dynamics
of the Holstein polaron, which they characterized by a single relaxation time $\tau$ and
showed it to be proportional to $h/g^2$ (in the notation of Eq.~\ref{eq:h:h-h}). To
identify the polaron they used a more complicated correlator
$\gamma(x)=\sum_{\bm{i}}\langle
\hat{n}_{\bm{i}}\ha^\dagger_{\bm{i}+\bm{x}}\ha_{\bm{i}+\bm{x}}\rangle$.  Chen, Zhao and
Tanimura applied the hierarchical equations of motion to \emph{exciton-phonon} coupled
systems and likewise found excitonic localization~\cite{chen_dynamics_2015}. This is a
very relevant scenario of the energy transfer in organic molecules and in photovoltaic
devices. Finally, Kloss \emph{et al.}~\cite{kloss_multiset_2019} were able to invistigate
larger 1D and 2D systems using the matrix product state (MPS) method. The objective was to
study the self-localized behavior in $e$-ph systems and to demostrate the computational
benefits offered by the tensor network states.  Here we demonstrate that GKBA+ODE approach
not only allows to study large systems of comparable spatial extent, but also to
incorporate $e$-$e$ interactions without imposing any restrictions on the number of
particles or the system dimensionality (see Tab.~\ref{tab:2} for a summary of systems and
computational resources).  This represents two great benefits of the GF methods in
contrast to other approaches: the dimension of the Hilbert space in the WF approach grows
factorially with the number of particles, whereas the MPS approach is more suited for 1D
systems.

Our numerical investigations are structured as follows. In the first step we establish
that for the treatment of the time-evolution of a correlated spin-0 pair of electrons the
$T$-matrix method in the $pp$ channel is the most appropriate. We demonstrate this by
comparing with the exact solution in the absence of phonons.  We will contrast very
similar exact and $T^{pp}$ results with the results from time-dependent Hartree-Fock
(TDHF) and from the second Born (2B) approximations. We then add $e$-ph interactions and
compare the localization of the electronic wave-packets at different $e$-$e$ interaction
strengths.

%    ---------- IV.A -----------
\section{Spreading of the two-electron wave-packet on the 1D lattice\label{sec:e-e}}
Here we consider the pure electronic case ($g=0$). The respective (Hubbard) Hamiltonian is
denoted as $\hat{H}_\text{h}$ (cf. Eq.~\ref{eq:h:h-h}). Electronic evolution starts from
the following initial condition
\begin{align}
  |\psi_{i}\rangle&\equiv|\psi(t=0)\rangle=\hd_{\bm{i}_0\uparrow}^\dagger \hd_{\bm{i}_0\downarrow}^\dagger|0\rangle,
  \label{eq:psi:0}
\end{align}
where $|0\rangle$ is the vacuum (empty) state, and $\bm{i}_0$ is the lattice site (typically in
the middle of the 1D chain) where the electrons are added.

The exact two-electron \emph{singlet} wave-function can be represented in a matrix form as
\begin{align}
  |\psi(t)\rangle&=\sum_{\bm{ij}}\frac{C_{\bm{ij}}(t)}{2}\left(\hd_{\bm{i}\uparrow}^\dagger \hd_{\bm{j}\downarrow}^\dagger
  -\hd_{\bm{i}\downarrow}^\dagger \hd_{\bm{j}\uparrow}^\dagger
  \right)|0\rangle,
\end{align}
with symmetry $C_{\bm{ij}}=C_{\bm{ji}}$. It is normalized as
$\langle\psi(t)|\psi(t)\rangle=\sum_{\bm{ij}}|C_{\bm{ij}}(t)|^2=1$ and fulfills the EOM
$i\frac{d}{dt}|\psi(t)\rangle=\hat{H}_\text{h}|\psi(t)\rangle$
\begin{multline}
  i\dot{C}_{\bm{ij}}(t)=-\left(C_{\bm{i}+\bm{1},\bm{j}}+C_{\bm{i}-\bm{1},\bm{j}}+C_{\bm{i},\bm{j}+\bm{1}}+C_{\bm{i},\bm{j}-\bm{1}}\right)\\
  +U\delta_{\bm{ij}}C_{\bm{ij}},
  \label{eq:TDSE}
\end{multline}
with a boundary condition that $C_{\bm{ij}}=0$ if $\bm{i},\bm{j}=0,\,N+1$, where $N$ is the number of
lattice sites.

Our main observables are the density matrix
\begin{subequations}
\begin{align}
  \rho^<_{\bm{ij},\sigma}(t)&\equiv \langle\psi(t)|\hd_{\bm{j}\sigma}^\dagger \hd_{\bm{i}\sigma}|\psi(t)\rangle=
  \sum_{\bm{n}} C^*_{\bm{in}}(t)C_{\bm{jn}}(t),
\end{align}
the electronic states' occupation numbers (independent of spin)
\begin{align}
  n_{\bm{i},\sigma}(t)&\equiv \langle\psi(t)|\hat{n}_{\bm{i}\sigma}|\psi(t)\rangle=\sum_{\bm{n}}|C_{\bm{in}}(t)|^2,\label{eq:ne}  
\end{align}
and the \emph{doublon} occupations 
\begin{multline}
  d_{\bm{i}}(t)\equiv \langle\psi(t)|(\hat{n}_{\bm{i}\uparrow}-n_{\bm{i}\uparrow})(\hat{n}_{\bm{i}\downarrow}-n_{\bm{i}\downarrow})|\psi(t)\rangle\\
  =|C_{\bm{ii}}(t)|^2- n_{\bm{i}\uparrow}n_{\bm{i}\downarrow}.
  \label{eq:doublon}
\end{multline}
\end{subequations}

Electron dynamics described by Eq.~\eqref{eq:TDSE} is not trivial and represent a
stringent test for approximations of MBPT. Even the dynamics of a \emph{single electron on
  a lattice} is quite different from the continous wave-packet spreading, which is
well-known from quantum mechanics. The difference comes from the form of the initial
state~\eqref{eq:psi:0}, which is spatially too narrow in comparison with the lattice
spacing as discussed by Sch\"{o}nhammer~\cite{schonhammer_unusual_2019}.

\begin{figure}[]
\centering  \includegraphics[width=0.95\columnwidth]{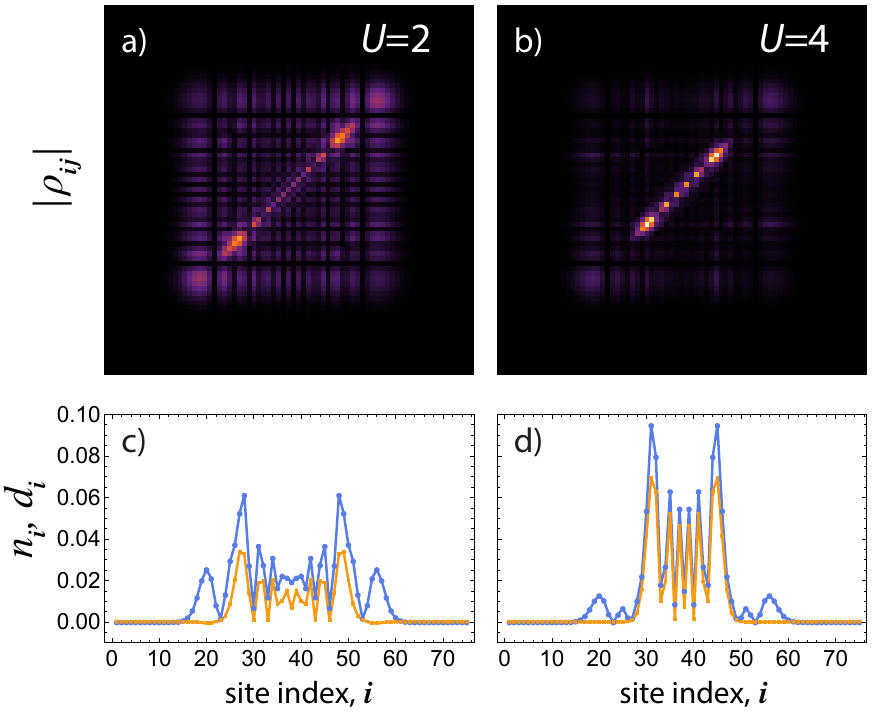}
\caption[]{Electronic properties of the 1D Hubbard chain of length $N=75$ comprising two
  electrons from the exact solution of Eq.~\eqref{eq:TDSE} with the initial condition
  $C_{ij}(0)=\delta_{ii_0}\delta_{ji_0}$ with $i_0=38$. Top row: snapshots at time $t=10$
  of the electronic density matrix. Bottom row: electron (blue) and doublon (orange)
  occupation numbers computed according to Eq.~\eqref{eq:ne} and Eq.~\eqref{eq:doublon},
  respectively.  \label{fig:1dhub:ex}}
\end{figure}

The \emph{electronic group velocity} $v_e$ is bounded:
\begin{align}
  v_e&=\left.\frac{d \epsilon(k)}{dk}\right|_{k=k_f}\le 2h.\label{eq:ve}
\end{align}
Here $\epsilon(k)= -2 h \cos(k)$ is the electron momentum dispersion relation and $k_f$ is
a center in momentum space of the fastest moving component of the wave-packet. When there
are more electrons in the system, the Hubbard interaction complicates the picture due to
the appearence of the resonant doublon
states~\cite{claro_interaction-induced_2003,rausch_filling-dependent_2017}. Since the
doublon dispersion differs from the electron dispersion by a prefactor: $E(k)= -J
\cos(k)$, where $J=4h^2/U$ (well justified for $U\gg h$) is the superexchange coupling
constant~\cite{rausch_filling-dependent_2017}, the doublon group velocity $v_d$ is in
general different from the electronic one.  This difference between the $v_e$ and $v_d$
can be appreciated by comparing blue and orange lines in Fig.~\ref{fig:1dhub:ex}.

\begin{figure}[]
\centering  \includegraphics[width=0.99\columnwidth]{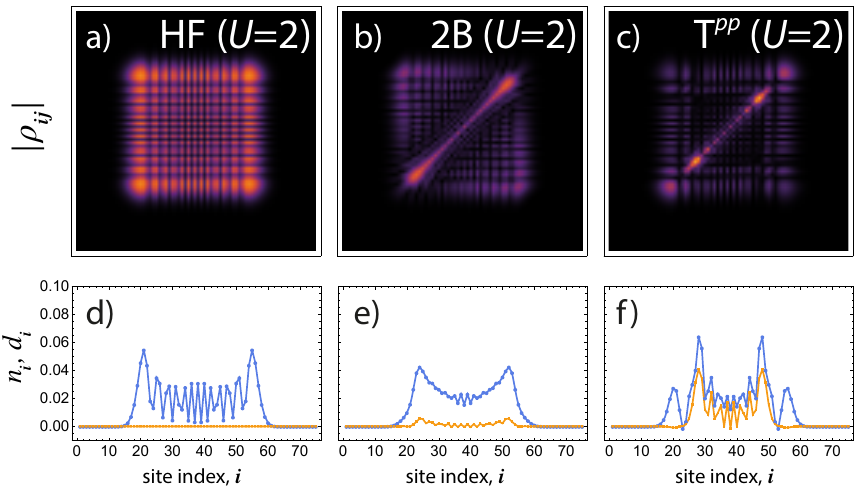}
\caption[]{Moderate Hubbard repulsion $U=2$: snapshots at time $t=10$ of the electronic density
  matrix (top row) and the state (blue lines) and doublon (orange lines) occupations
  (bottom row) obtained using GKBA+ODE with different
  approximations.  \label{fig:1dhub:u2}}
\end{figure}
\begin{figure}[]
\centering  \includegraphics[width=0.99\columnwidth]{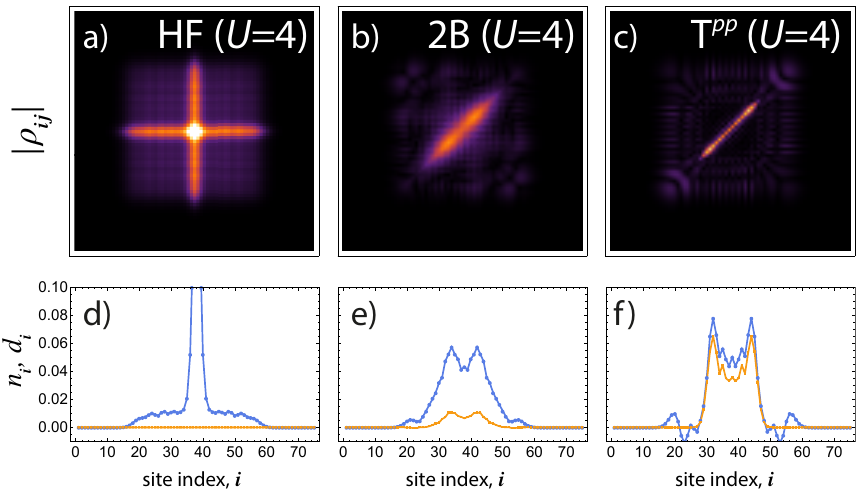}
\caption[]{Strong Hubbard repulsion $U=4$: snapshots at time $t=10$ of the electronic
  density matrix (top row) and the state (blue lines) and doublon (orange lines)
  occupations (bottom row) obtained using GKBA+ODE with different
  approximations.  \label{fig:1dhub:u4}}
\end{figure}

In Figs.~\ref{fig:1dhub:u2},\,\ref{fig:1dhub:u4}, the performance of three different
methods: time-dependent Hartree-Fock (HF), second Born approximation and the $T$-matrix
approximation in the $pp$ channel is compared against these exact results. The initial
condition for the electronic density matrix corresponding to the initial state
Eq.~\eqref{eq:psi:0} reads
\begin{align}
  \rho_{\bm{ij},\sigma}(t=0)&=\delta_{\bm{ii}_0}\delta_{\bm{ji}_0},&
   \mcG^{e}(t=0)&=0,\label{eq:rho:0}
\end{align}
meaning that the two electrons (with spin $\uparrow$ and $\downarrow$) are
initially uncorrelated.

\begin{figure}[h!]
\centering  \includegraphics[width=\columnwidth]{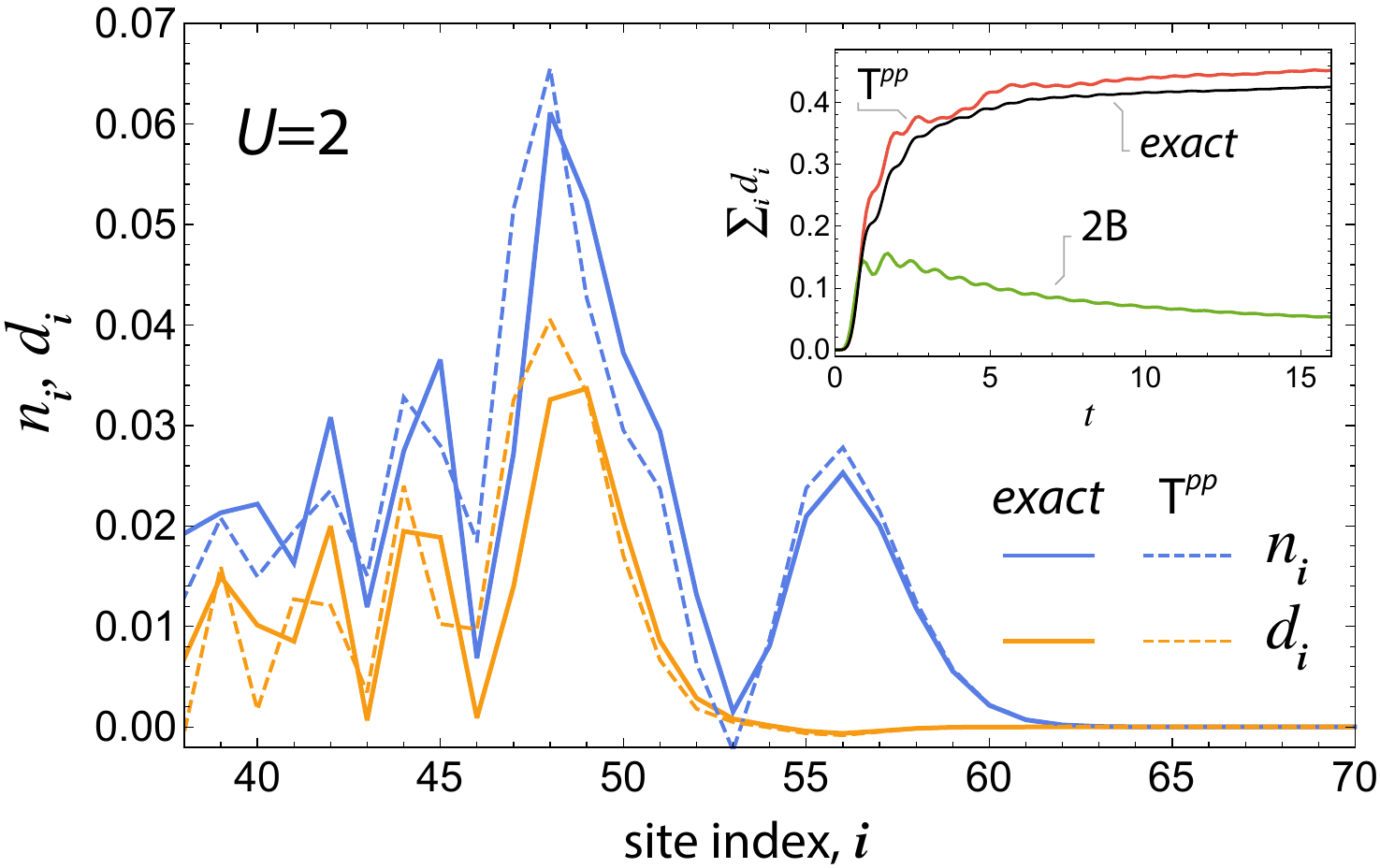}
\caption[]{Comparison of the electron $n_{\bm{i}}\equiv n_{\bm{i},\sigma}$ (blue) and
  doublon occupations $d_{\bm{i}}$ (orange) for a Hubbard chain with $N=75$ at time $t=10$
  illustrating the difference in the corresponding group velocities.  Due to the inversion
  symmetry only a half of the system is shown. The inset depicts the total number of
  doublons in the system from the exact time-evolution vs. 2B and $T^{pp}$
  approximations.\label{fig:1dhub:cf:ex}}
\end{figure}

As expected, the mean-field method is completely inadequate for the 1D Hubbard model
considered here: the electronic density matrix features a four-fold symmetry pertinent to
noninteracting particles. A bit more accurate is the 2B method. It correctly predicts the
density localization at the diagonal, however, it fails to describe the spreading of
doublons: at $U=2$ they propagate with nearly the same velocity as electrons
(Fig.~\ref{fig:1dhub:u2}). The situation seems to be different from the half-filled case,
where already 2B approximation is accurate for the prediction of
doublons~\cite{balzer_doublon_2018}. Only $T^{pp}$ provides good agreement with the exact
solution for the two-electron system considered here for intermediate ($U=2$) and large
($U=4$) Hubbard repulsion. This can be seen in the density matrix plots in
Figs.~\ref{fig:1dhub:u2},\,\ref{fig:1dhub:u4}, as well as in Fig.~\ref{fig:1dhub:cf:ex},
where we zoom in into the electron and doublon occupation numbers as functions of the
lattice site. These two observables can be directly obtained from the GKBA+ODE
approach. The first one, the electron occupation number is given by
\begin{align}
n_{\bm{i},\sigma}(t)&=\rho^<_{\bm{ii},\sigma}(t).
\end{align}
For spin-compensated systems as considered here,
$\rho_{\bm{ij},\uparrow}=\rho_{\bm{ij},\downarrow}$, which allows to drop the spin index, i.e.,
$n_{\bm{i}}\equiv n_{\bm{i},\sigma}$.

The second observable, the number of doublons $d_{\bm{i}}(t)$ as a function of the site
number $\bm{i}$, can be considered in parallel to the wave-function approach presented
above. Puig von Friesen, Verdozzi and Almbladh demonstrated how this quantity can be
computed in the full Kadanoff-Baym method applied to the $T$-matrix
approximation~\cite{puig_von_friesen_can_2011}. In the GKBA+ODE scheme the doublon
occupations are expressed in terms of the two-body correlator $\mcG^{e}(t)$
\begin{align}
  d_{\bm{i}}(t)&=\langle\psi(t)|\hat{n}_{\bm{i}\uparrow}\hat{n}_{\bm{i}\downarrow}|\psi(t)\rangle
  -n_{\bm{i}\uparrow}(t)n_{\bm{i}\downarrow}(t)=-\cG^{e}_{\bm{iiii}}(t).
\end{align}
Notice that we are talking here about the correlated part of the doublon occupation number
as defined by Eq.~\eqref{eq:doublon}. Not only Fig.~\ref{fig:1dhub:cf:ex} demonstrates
excellent agreement between the exact and the $T^{pp}$-matrix methods, it also nicely
illustrates the difference between two different group velocities $v_e$ and $v_d$.

Notice that on physical grounds we do not expect similar performance from $GW$ and
$T^{ph}$ methods (they reflect correlations in particle-hole channels, which are not
relevant here). Therefore, all subsequent results with phonons will be presented only for
these three methods: HF, 2B and $T^{pp}$. 2B approximation is accurate only for initial
instances of time. At larger times the population of doublons according to the 2B
approximation diminishes, in contrast with the prediction of the $T^{pp}$ approximation,
which is in good agreement with exact results (see inset of
Fig.~\ref{fig:1dhub:cf:ex}). Spectral properties of two-electron systems have also been
studied with similar conclusions: $T^{pp}$ is very accurate, whereas 2B approximation
underestimates the binding energy of a Cooper pair (see Fig.\,13.5 in
Ref.~\cite{stefanucci_nonequilibrium_2013}).

%    ---------- IV.B -----------
\section{Role of phonons: Ehrenfest vs. $GD$ approximation \label{sec:e-ph}}
\begin{figure*}[]
\centering  \includegraphics[width=1.55\columnwidth]{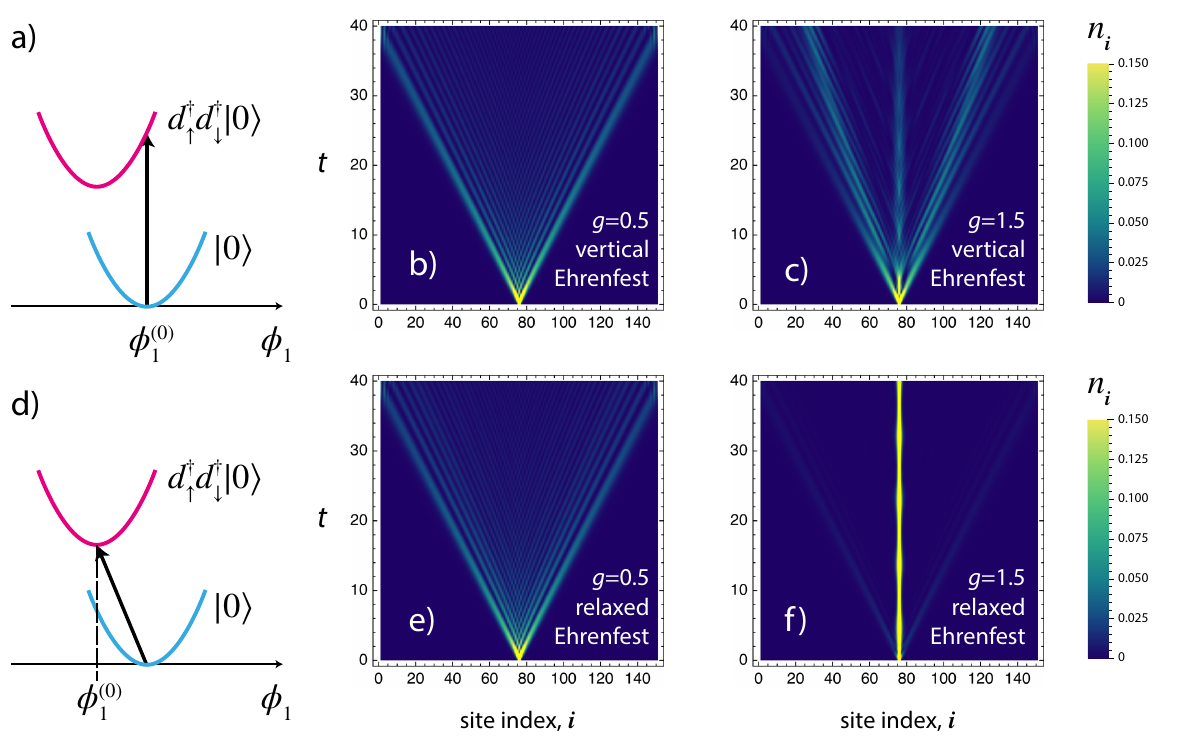}
\caption[]{Evolution of the electronic density $n_{\bm{i}}\equiv n_{\bm{i},\sigma}$ for
  different $e$-ph interaction strengths and initial states as a function of time ($t$)
  and lattice site $\bm{i}$. Phonons are treated semiclassically, i.\,e., in the Ehrenfest
  approximation. Top row depicts the case of a ``vertical'' initial state~\eqref{eq:vert}
  for phonons, bottom row\,---\,the ``relaxed'' initial state~\eqref{eq:relax}.  In panels
  (a) and (d) these two scenarios are schematically depicted for the $\bm{i}_0$ lattice
  site. Panels (b) and (e) the negligible effect of the initial state for the case of weak
  $e$-ph interaction ($g=0.5$). Panels (c) and (f) demonstrate that for the strong $e$-ph
  interaction ($g=1.5$) the electronic density is localized at the central site when the
  dynamics is started from the relaxed phononic state. Initial electronic density matrix
  is the same in all cases and is given by Eq.~\eqref{eq:rho:0}.
  \label{fig:ehrenfest}}
\end{figure*}

While the Holstein model is not solvable for arbitrary parameters, there are regimes
where the system behavior can be understood at least qualitatively. If the mass of a
phonon is much larger than the mass of an electron, the system is in the \emph{adiabatic
  regime}. It is characterized by the well-defined potential energy surfaces and large
stiffness of the lattice and is attained for $\omega/h\ll1$. In the opposite
anti-adiabatic limit the phonons adjusts almost instantaneously to the motion of an
electron. We consider here the more intriguing cross-over regime $\omega/h=1$.

The binding energy of a polaron can be estimated
perturbatively~\cite{langreth_singularities_1970} as
\begin{align}
  \varepsilon_p&=-\frac{g^2}{\omega}.
\end{align}
By comparing it with the electronic band-width of $2h$ we arrive at the second important
ratio:
\begin{align}
  \lambda&=\frac{g^2}{2h \omega}.
\end{align}
When $\lambda\ll1$, electrons in a small portion of the Brillouin zone around the band
bottom are affected by phonons, lattice distortions spread over many lattice sites, and
one classifies this quasiparticle as the ``large'' or ``light'' polaron. The polaron
effective mass is enhanced as compared to the mass of an electron, however, the mass
renormalization depends only on the phononic frequency, and not on the interaction
strength. We will see below that this is not the case for our simulations. In the opposite
limit of the so-called ``small'' or ``heavy'' polaron all electronic states are dressed by
phonon~\cite{devreese_frohlich_2009}. A large exponential renormalization of the hopping
constant is a marked physical phenomenon in this
limit~\cite{stefanucci_nonequilibrium_2013}:
\begin{align}
  \tilde{h}&=h e^{-g^2/\omega^2}.\label{eq:h:renorm}
\end{align}
Here we consider two values $e$-ph coupling $g=0.5$ and $1.5$ corresponding to
$\lambda=0.125$ and $1.125$, respectively.

We start the investigation of the $e$-ph dynamics with the simplest semiclassical
treatment of the optical phonons and without $e$-$e$ interactions ($U=0$). This is the
scenario of Kloss \emph{et al.}~\cite{kloss_multiset_2019}, earlier investigations are by
Sayyad and Eckstein~\cite{sayyad_coexistence_2015} (initially hot electron distribution,
DMFT), Dorfner \emph{et al.}~\cite{dorfner_real-time_2015} (diagonalization in a limited
functional space). While the methods are very different, in our approach we can treat
systems of comparative sizes (we use $N=151$ for all subsequent results) and even with
$e$-$e$ interactions (next section). Since now the phononic subsystem is included, the
initial conditions for the electrons are supplemented with the conditions for the phononic
density matrix $\mgamma^<$, the electron-phonon two-body correlator $\mcG^\text{ph}$ and the
phononic coordinates $\phi$
\begin{subequations}
\begin{align}
  \gamma^<_{\bm{i}\xi,\bm{j}\zeta}(t=0)&=\delta_{\bm{ii}_0}\delta_{\bm{ji}_0} \gamma^{(0)}_{\xi\zeta},
  &\gamma^{(0)}&=\frac12\begin{pmatrix}
      1  & -i \\
      i & 1
  \end{pmatrix};\label{ic:gamma}\\
  \mcG^\text{ph}(t=0)&=0;\label{ic:Gb}\\
  \phi_{\bm{i},\xi}(t=0)&=\delta_{\bm{ii}_0}\phi^{(0)}_\xi.
\end{align}
\end{subequations}
Eq.~\eqref{ic:gamma} indicates that the initial phononic state is a coherent state
\begin{align}
  |\varphi_i\rangle&\equiv|\varphi(t=0)\rangle=e^{\beta \ha_{\bm{i}_0}^\dagger- \beta^*
    \ha_{\bm{i}_0}}|0\rangle.
\end{align}
Eq.~\eqref{ic:Gb} implies that there are no $e$-ph correlations initially. In parallel to
Kloss \emph{et al.}~\cite{kloss_multiset_2019}, two scenarios for the initial phononic
displacements $\phi^{(0)}_\xi$ are considered. In the first one,
Fig.~\ref{fig:ehrenfest}(a), a \emph{vertical} transition takes place from the vacuum
state with zero number of electrons and phonons to a state with two electrons. In this
case the phonons retain their averaged values of the displacement and momentum:
\begin{align}
  \mphi^{(0)}&=\langle \varphi_i|\hphi_{\bm{i}_0}|\varphi_i\rangle=0,&
  \beta&=0.\label{eq:vert}
\end{align}
In the second \emph{relaxed} scenario, Fig.~\ref{fig:ehrenfest}(b), it is assumed that the
phonon coupled to the excitation site $\bm{i}_0$ is instantly displaced to a new minimum
of the potential energy surface:
\begin{align}
  \mphi^{(0)}&=\langle \varphi_i|\hphi_{\bm{i}_0}|\varphi_i\rangle=\begin{pmatrix}
  -g/\omega\\0
  \end{pmatrix},&
  \beta&=-g/\left(\sqrt{2}\omega\right).\label{eq:relax}
\end{align}

Time-evolution of the electronic occupations in the presence of classical phonons, i.\,e.,
Ehrenfest approximation ($\mgamma^<(t)=\mgamma^<(0)$, only Eq.~\eqref{eq:phi:eom} for
$\phi_\mu$ is solved for phonons), is depicted in Fig.~\ref{fig:ehrenfest}(central and
right columns). For $g=0.5$ [panels (b) and (e)], the electron wave-packet spreads without
any localization, and the effect of initial state on the electron dynamics is
negligible. For stronger $e$-ph interaction [$g=1.5$ panels (c) and (f)] the effect of
initial phononic state is strong. One observes an almost complete localization starting
from the relaxed configuration, which is rather unphysical. In particular, it would be
impossible to explain the localization by the phononic renormalization of the hopping
parameter.

\begin{figure}[]
\centering  \includegraphics[width=0.95\columnwidth]{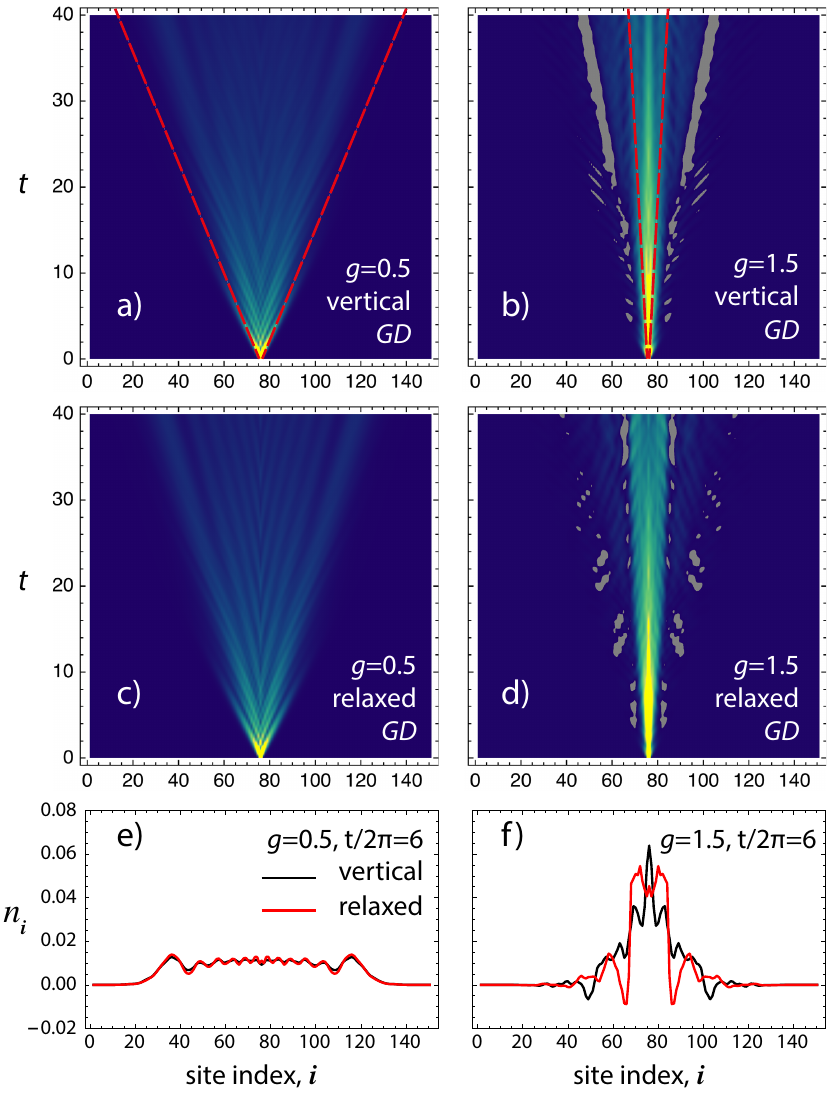}
\caption[]{(a-d) Evolution of the electronic density of the $N=151$ Holstein-Hubbard chain
  for different $e$-ph interaction strengths and initial states. Dashed lines in panels
  (a) and (b) schematically depict light-cones corresponding to the renormalized group
  velocity $v_{e}e^{-g^2/\omega^2}$. (e-f) electronic densities at time
  $t/(2\pi)=6$. $e$-ph correlations are treated on the $GD$ level. \label{fig:GD}}
\end{figure}

However, if we add the $e$-ph correlation effects at the $GD$ level (Fig.~\ref{fig:GD}), a
dramatic improvement in the spatial extent of the electron density can be seen in close
agreement with the finding of Ref.~\cite{kloss_multiset_2019} that ``the Franck-Condon
excitation is seen to retain a substantial mobility even under strong coupling''. The
electron group velocity reduction is well described by Eq.~\eqref{eq:h:renorm} as
indicated by red dashed lines.

We also notice that GKBA+ODE approach may sometimes lead to negative electronic densities
(Fig.~\ref{fig:1dhub:u4}(f) and grey areas in Fig.~\ref{fig:GD}) even though the total number of particles is
conserved. The weight of these domains is typically very small and only causes numerical
instabilities when the $e$-ph coupling constant is large. We mention in passing that
negative populations have been observed also in NEGF simulations with memory
truncation~\cite{schuler_truncating_2018}.

\begin{figure}[]
\centering  \includegraphics[width=0.95\columnwidth]{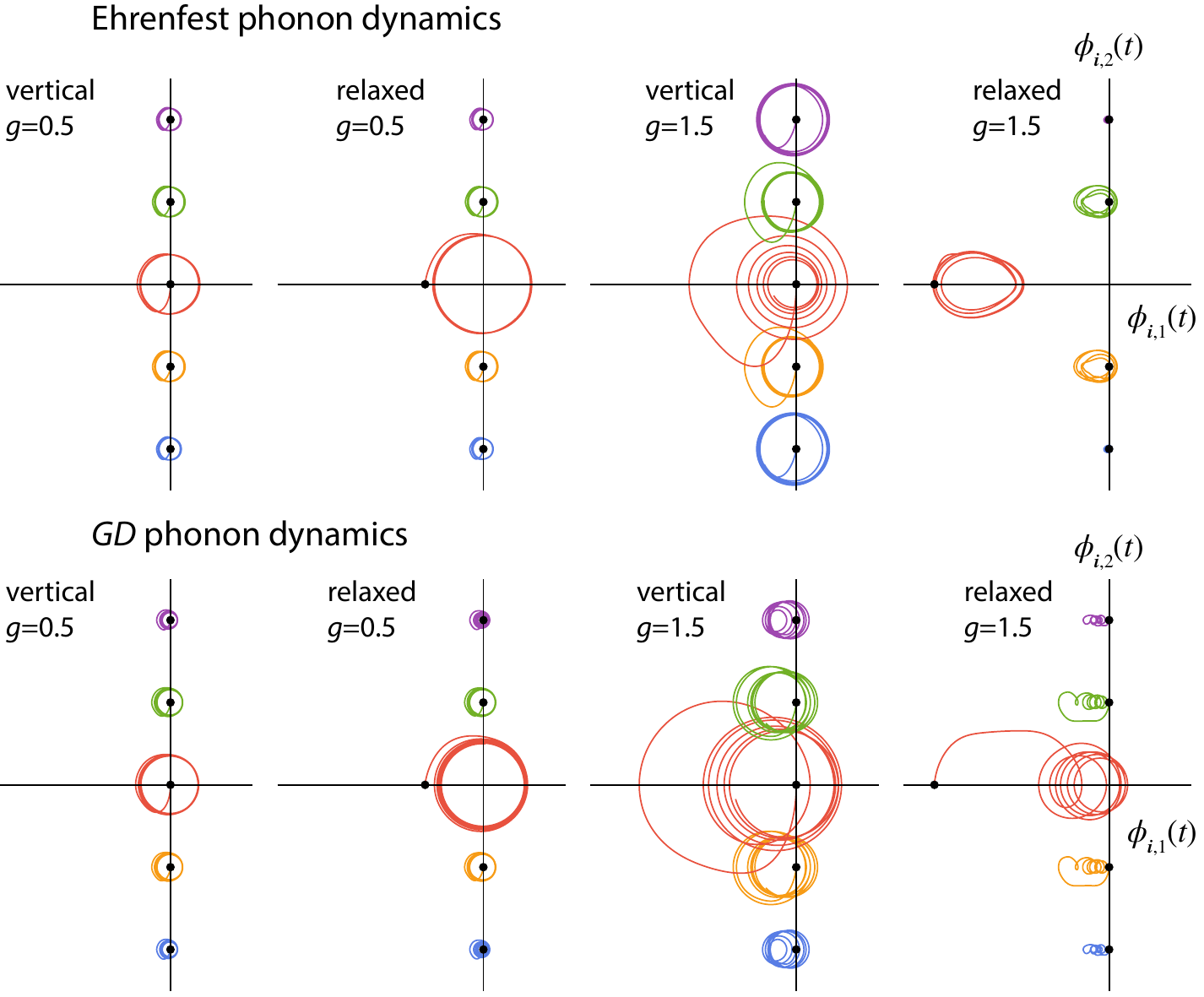}
\caption[]{Phase portrait of the five central phonons of the $N=151$ Holstein-Hubbard
  chain. Ehrenfest (top) and $GD$ (bottom) results are compared different $e$-ph
  interactions and initial states.\label{fig:phase}}
\end{figure}

Next, we look at the phononic observables, starting with the site-resolved averaged values
of the displacement and momentum, Fig.~\ref{fig:phase}. Our main finding here is that the
phase trajectories approach limiting cycles at the Ehrenfest level (in line with
Ref.~\cite{kartsev_nonadiabatic_2014}), whereas they are not closed at the $GD$ level
(this difference is more pronounced for $g=1.5$) indicating a damping of the phononic
subsystem. This important physical phenomenon has also been studied in the context of
polaron relaxation~\cite{golez_relaxation_2012}. We will revisit this issue in the next
section when discussing the total energy conservation.

%    ---------- IV.C -----------
\section{Role of $e$-$e$ interactions and phonons \label{sec:both}}
\begin{figure*}[t]
  \includegraphics[width=\textwidth]{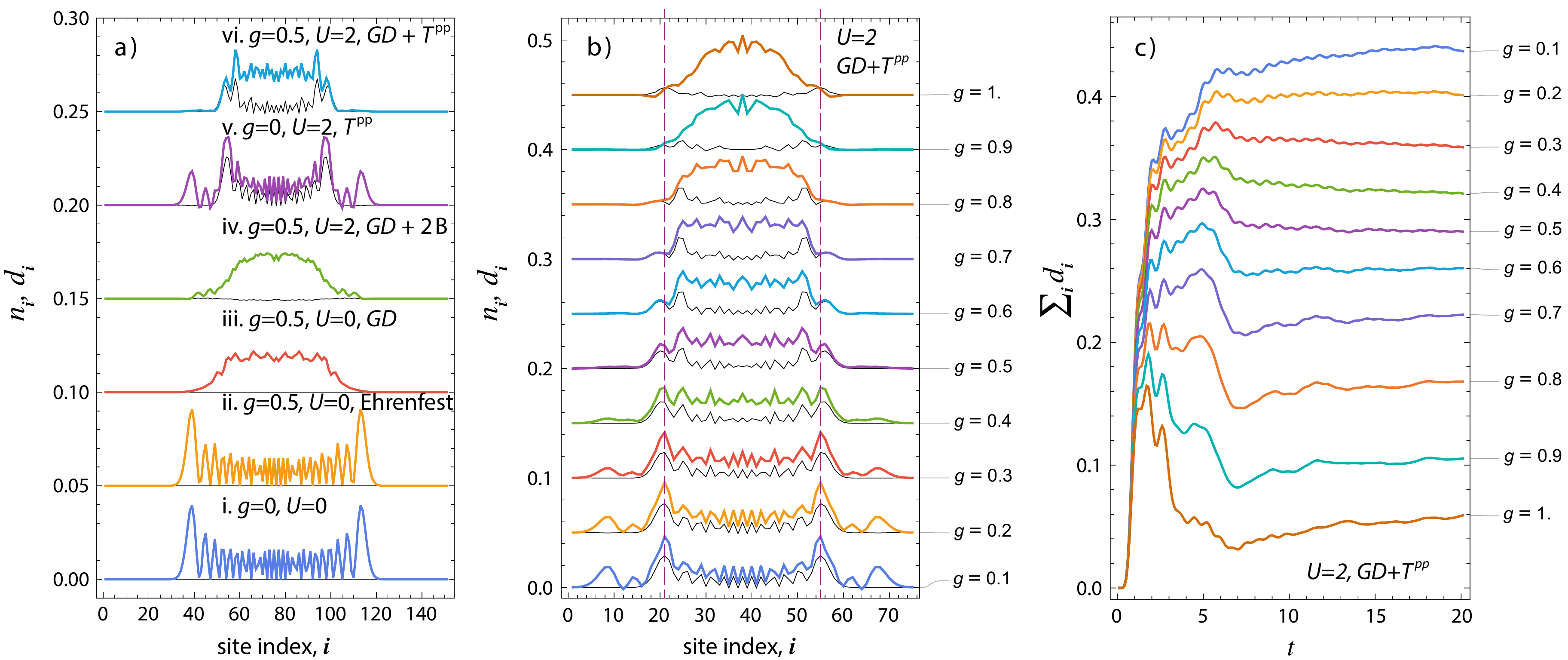}
\caption[]{(a) Electron occupations (color) and doublon occupations (thin black) at time
  $t=20$ for all used methods and a vertical initial state. (b) Snapshots at $t=16$ of the
  electronic occupations (color) and doublon occupations for Holstein-Hubbard model with
  $N=75$ for different $e$-ph interaction strengths. (c) Time-evolution of the total
  number of doublons in the system for different $e$-ph interaction
  strengths.\label{fig:scan:u2}}
\end{figure*}
In Sec.~\ref{sec:e-e} we have demonstrated that electron dynamics in the presence of
$e$-$e$ interactions is modified by the formation and propagation of doublon
excitations. Electron density matrix is fragmented into a diagonal part hosting doublons
and off-diagonal part describing singly-occupied lattice sites. In Sec.~\ref{sec:e-ph} we
have considered the effect of phonons on the non-interacting electrons and have shown how
it reduces their group velocity leading to self-trapping.  In Fig.~\ref{fig:scan:u2}(a) we
summarize the results of different methods with and without $e$-$e$ and $e$-ph
interactions.  It can be seen that they have very different effects on the system dynamics
(cf. subpanels v and iii).

It is well known that phonons hybridize with a variety of electronic excitations:
plasmons~\cite{huber_femtosecond_2005,dai_graphene_2015,verdi_origin_2017},
excitons~\cite{chen_dynamics_2015,li_excitonphonon_2021,stefanucci_carriers_2021},
polaritons~\cite{wu_topological_2015}, collective amplitude modes in
superconductors~\cite{murakami_multiple_2016}. Much less is known about the coupling of
doublons and phonons, being a topic of current intense
investigations~\cite{butler_doublonlike_2021}. The effect has been studied almost
exclusively in the half-filled case, where it was shown that phonons hybridize and modify
the doublon dispersion relation~\cite{seibold_time-dependent_2014}, the phonon-assisted
decay of excess doublons and the phonon-enhanced doublon production was
predicted~\cite{werner_phonon-enhanced_2013}. But what happens in the \emph{two-electron}
case?

In order to elucidate the simultaneous effect of $e$-$e$ and $e$-ph interactions, we fixed
$U=2$ and performed a series of calculations for various electron-phonon interaction
strengths, Fig.~\ref{fig:scan:u2}(b). Here we clearly see the already discussed
self-trapping of electrons. It is reflected in the width reduction of the electronic
distributions (depicted in color). Quite remarkably, however, the effect of the $e$-ph
interaction on the distribution of doublons (thin black lines) is \emph{minimal}. To
facilitate the comparison, vertical dashed lines in Fig.~\ref{fig:scan:u2}(b) mark the
position of the fastest doublon peaks for $g=0.1$. Increasing $e$-ph interaction from
$0.1$ to $0.7$, a slight increase of the width of doublon distribution becomes evident, a
part of the doublon density can be found outside the marked domain. By increasing $g$
further, the doublon distribution is squeezed again, whereby the peak position for $g=1$
coincides with the vertical line. This non-monotonous behaviour indicates that two
competing physical phenomena are at play. A qualitative understanding is provided by the
superexchange expression for the doublon hopping constant $J=4h^2/U$, which is the measure
of the group velocity $v_d$. $e$-ph interaction modifies both parameters entering $J$. The
Hubbard-$U$ constant is renormalized due to the emergence of the effective $e$-$e$
interaction mediated by phonons~\cite{van_leeuwen_first-principles_2004}. From general
arguments, the effect is positive and nonlocal in time~\cite{hirsch_phase_1983}. In the
extreme antiadiabatic regime the interaction becomes instantaneous leading to an effective
Hubbard model with a reduced $U$ value~\cite{de_filippis_quantum_2012}\,---\,hence the
enhancement of $v_d$.  However, for the cross-over case $\omega=h$ considered here, the
effect is small, and therefore for larger $g$ the exponential renormalization of the
hopping constant according to Eq.~\eqref{eq:h:renorm} becomes the dominant mechanism. It
leads to the slow-down of the doublon spreading seen for $g\ge0.8$.

In Fig.~\ref{fig:scan:u2}(c), the total number of doublons in the system is plotted as a
function of time for different $e$-ph coupling strengths. It features a rapid increase of
the population at initial stage, followed by a depopulation (for $g\ge0.3$) after which a
steady value is reached in an oscillatory fashion on a much longer time scale. In line
with the discussion above, the steady number of doublons is a monotonically decreasing
function of $g$ confirming again the renormalization of the Hubbard-$U$ by $e$-ph
interactions.

\begin{figure}[b!]
\centering  \includegraphics[width=\columnwidth]{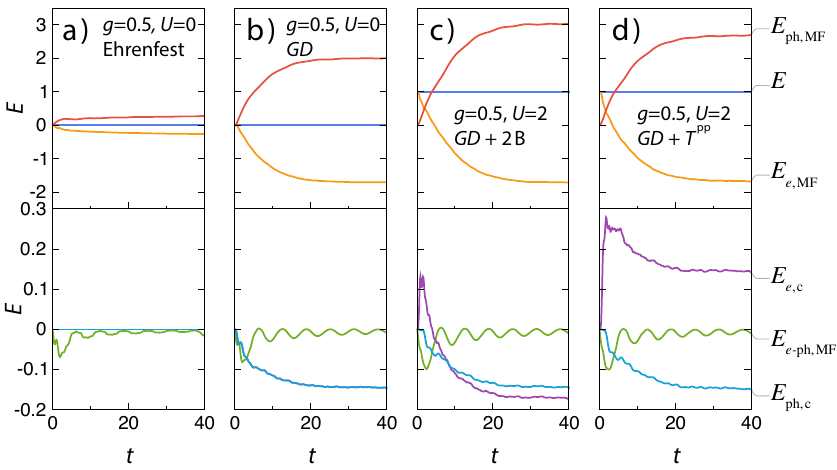}
\caption[]{Time-dependent energies for the $g=0.5$ cases with vertical initial state.\label{fig:eng:g05}}
\end{figure}

\begin{figure}[h!]
\centering  \includegraphics[width=\columnwidth]{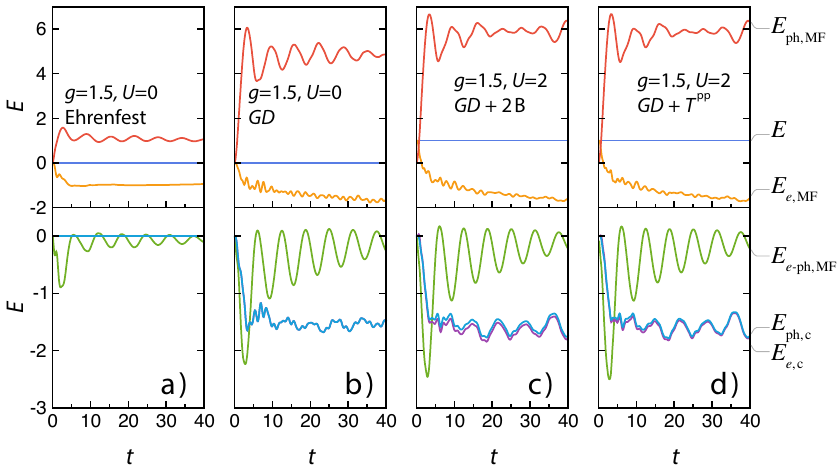}
\caption[]{Time-dependent energies for the $g=1.5$ cases with vertical initial state.\label{fig:eng:g15}}
\end{figure}

The discussion of the doublon dynamics is facilitated by the analysis of the total energy
and its various contributions:
\begin{align}
  E(t) = E_{e,\,\text{MF}}(t) + E_{\text{ph,\,MF}}(t)+ E_\text{$e$-ph,\,MF}(t) + E_\text{c}(t).
\end{align}
The mean-field energy contributions are defined as follows:
\begin{subequations}
\begin{align}
  E_{e,\,\text{MF}}(t) &= \frac12\Tr [(h(t)+h_\text{HF}) \rho(t)],\label{eq:E:e:MF}\\
  E_{\text{ph,\,MF}}(t) &= \Tr [\mO(t) \mGamma(t)], \label{eq:E:b:MF}\\
  E_\text{$e$-ph}(t) &= \Tr [h^\text{$e$-ph}(t) \rho(t)],\label{eq:E:e-b}
\end{align}
\end{subequations}
with $\Gamma_{\mu\nu}(t)=\gamma_{\mu\nu}^{<}(t)+\phi_{\mu}(t)\phi_{\nu}(t)$ being the full
phononic density matrix. The correlation energy $E_\text{c}$ is associated with $\mcG^{e}$
and $\mcG^\text{ph}$
\begin{align}
E_\text{c}&=E_{e,\,\text{c}}+E_{\text{ph,\,c}}.\label{eq:E:c}
\end{align}
Explicit expressions can be formulated on the Keldysh contour:
\begin{subequations}
 \begin{align}
   E_{e,\,\text{c}}(t)&=-\frac{i}{2}\int_\gamma d\bar{t}\,\Tr\Sigma^e(t,\bar{t})G(\bar{t},t^+),\\
   E_{\text{ph,\,c}}(t)&= \frac{i}{2}\int_\gamma d\bar{t}\,\Tr\mSgm^\text{ph}(t,\bar{t})\mD(\bar{t},t^+),
 \end{align}
\end{subequations}
where $\Sigma^e$ and $\mSgm^\text{ph}$ is electronic, phononic self-energy, respectively.
For conserving approximations they can be written in a symmetrized form in terms of the
collision integrals:
\begin{subequations}
\begin{align}
E_{e,\,\text{c}}(t) &= -\frac{i}{4} \Tr \big[ I^e(t) -  \left(I^e(t)\right)^\dagger\big], \\
E_{\text{ph,\,c}}(t)  &= \frac{i}{4} \Tr \big[ \ma \mI^\text{ph}(t) - \left(\ma \mI^\text{ph}(t)\right)^\dagger\big].
\end{align}
\end{subequations}
The two contributions are equal in the absence of $e$-$e$ interactions as has been shown
in our earlier work~\cite{karlsson_fast_2021}.  Notice that $E_{e,\,\text{c}}$ contains a
phononic part (equal to $E_{\text{ph\,c}}$) and a pure $e$-$e$ correlation energy. In
Figs.~\ref{fig:eng:g05},\ref{fig:eng:g15}, the time-evolution of the total energy
ingredients is presented for two values of $e$-ph coupling strength, $g=0.5$ and $g=1.5$
and two values of $e$-$e$ coupling, $U=0$ and $U=2$. In all cases the total energy is
conserved up to numerical accuracy despite the fact that correlated contributions are an
order of magnitude smaller. Theories with \emph{frozen} phonons obviously do not fulfill
this property.

The initial stage of the doublon dynamics is driven by a pure electronic mechanism. This
can be concluded by comparing a very quick increase of the electronic correlational energy
$E_{e,\,\text{c}}$ (proportional to the number of doublons) and a slow variation of the
mean-field phononic energy $E_{\text{ph,\,MF}}(t)$ (proportional to the number of phonons) in
Figs.~\ref{fig:eng:g05} ($GD$+$T^{pp}$ results). The latter grows in time at the expense
of the electron kinetic energy $E_{e,\,\text{MF}}$~\cite{schuler_truncating_2018}.  In
panels (a) to (b) in Figs.~\ref{fig:eng:g05} and \ref{fig:eng:g15} the kinetic energy
exchange between the electronic and phononic subsystems is severely underestimated. This is
a shortcoming of the Ehrenfest approximation and it is cured in $GD$. Notice that the
efficiency of the kinetic energy exchange increases with increasing $g$ and remains
unchanged even in the presence of $e$-$e$ interaction.

By inspecting the behavior of the pure electronic correlation energy $E_{\text{$e$-$e$,
    c}}=E_{e,\,\text{c}}-E_{\text{ph,\,c}}$ in the asymptotic regime we conclude that for
larger times the doublon-phonon scattering becomes less efficient; the energy of doublons
is insufficient to excite phonon quanta\,---\,the so-called phononic bottleneck effect.
In 2B pure electronic correlations given by $E_{\text{$e$-$e$, c}}$ is negative
(difference between magenta and cyan lines), meaning that the density of doublons goes
negative, see also panel (iv) in Fig.~\ref{fig:scan:u2}(a). This shortcoming of the 2B is
cured by the $T^{pp}$ approximation.

\begin{figure}[th!]
\centering  \includegraphics[width=\columnwidth]{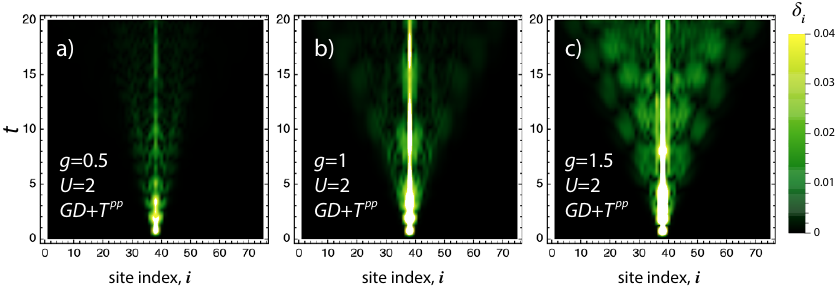}
\caption[]{The $e$-ph correlator $\delta_{\bm{i}}=\sqrt{\delta_{\bm{i},1}^2+\delta_{\bm{i},2}^2}$ as a
  function of time for three different $e$-ph coupling constants and vertical initial
  state.\label{fig:geeb}}
\end{figure}

Finally we characterize the spreading of the polaronic quasiparticle. Its spatial
extension can be quantified by plotting the correlators
\begin{align}
  \delta_{\bm{i},\xi}(t)&=\left\langle \hat{n}_{\bm{i}_0}(t)\hphi_{\bm{i}_0+\bm{i},\xi}(t)\right\rangle
  =\cG^\text{ph}_{\bm{i}_0+\bm{i},\xi;\bm{i}_0\bm{i}_0}(t),  \label{eq:polaron:corr}
\end{align}
where $\bm{i}_0$ is the initial excitation site. It represents a conditional probability
of a phonon at site $\bm{i}_0+\bm{i}$ having the mean displacement $\phi$ when an electron
is at site $\bm{i}_0$. Therefore, it can be used as a tool to distinguish the coherent
phononic cloud propagating together with the electron from uncorrelated phononic
background~\cite{fehske_spatiotemporal_2011}. As we already discussed above, our system in
the presence of strong $e$-$ph$ interaction, behaves in many ways similar to the case with
reduced $e$-$e$ coupling. This can be seen in the dynamics of doublons in
Fig.~\ref{fig:scan:u2}(b), and in similarities between panels (b) and (d) in
Fig.~\ref{fig:eng:g15}. Therefore, Fig.~\ref{fig:geeb}(c) depicts the classical picture of
polaron spreading. Going to smaller values of $e$-$ph$ coupling as in
Fig.~\ref{fig:geeb}(a,\,b), the effect of $U$ becomes dominant, it reduces the spectral
strength and spatial extent of a polaron, representing a marked feature of electronic
correlations.

\section{Summary\label{sec:summary}}
In this work, we applied a nonequilibrium Green's function formalism (paper I) to
investigate coupled electron-phonon dynamics in 1D Hubbard-Holstein model. While theories
separately describing $e$-$e$ correlations or $e$-ph correlations in model systems are
known, this contribution is devoted to the systems where both ingredients are
important. We restrict to the $\Phi$-derivable and, therefore, conserving approximations,
and consider a class of theories with an additive Baym functional consisting of the
electronic and phononic parts. This leads to intertwined electronic and phononic
self-energies, whereas the electronic and phononic response functions are treated
independently. This approach captures with elegance the feedback in time-domain of phonons
on the electronic properties and the modification of phononic properties caused by the
dynamics of electrons. The method scales linearly with the physical propagation time
thanks to the use of GKBA in the ODE formulation. This allows one to perform simulations
for unprecedentedly large systems such as 1D chains described by the extended
Holstein-Hubbard model. The model is a prototype for typical photovoltaic systems and
possesses two remarkable quasiparticle states\,---\,polarons and doublons. The interplay
between them leads to intricate physical phenomenon of the doublon localization that we
discuss at length here. We emphasize that the phenomenon is manifested at moderate $e$-$e$
and $e$-ph interaction strengths and therefore is well suited for methods based on
many-body perturbation theory. Besides the obvious observables such as electronic
occupation numbers and density matrices, phase-space dynamics of individual phonons, we
consider more complicated $e$-ph correlators and ingredients of the total energy.

\begin{acknowledgments}
  We acknowledge the financial support from MIUR PRIN (Grant No. 20173B72NB), from INFN
  through the TIME2QUEST project, and from Tor Vergata University through the Beyond
  Borders Project ULEXIEX.
\end{acknowledgments}

%\bibliography{MyLibrary,Amendments}

\begin{thebibliography}{63}%
\makeatletter
\providecommand \@ifxundefined [1]{%
 \@ifx{#1\undefined}
}%
\providecommand \@ifnum [1]{%
 \ifnum #1\expandafter \@firstoftwo
 \else \expandafter \@secondoftwo
 \fi
}%
\providecommand \@ifx [1]{%
 \ifx #1\expandafter \@firstoftwo
 \else \expandafter \@secondoftwo
 \fi
}%
\providecommand \natexlab [1]{#1}%
\providecommand \enquote  [1]{``#1''}%
\providecommand \bibnamefont  [1]{#1}%
\providecommand \bibfnamefont [1]{#1}%
\providecommand \citenamefont [1]{#1}%
\providecommand \href@noop [0]{\@secondoftwo}%
\providecommand \href [0]{\begingroup \@sanitize@url \@href}%
\providecommand \@href[1]{\@@startlink{#1}\@@href}%
\providecommand \@@href[1]{\endgroup#1\@@endlink}%
\providecommand \@sanitize@url [0]{\catcode `\\12\catcode `\$12\catcode
  `\&12\catcode `\#12\catcode `\^12\catcode `\_12\catcode `\%12\relax}%
\providecommand \@@startlink[1]{}%
\providecommand \@@endlink[0]{}%
\providecommand \url  [0]{\begingroup\@sanitize@url \@url }%
\providecommand \@url [1]{\endgroup\@href {#1}{\urlprefix }}%
\providecommand \urlprefix  [0]{URL }%
\providecommand \Eprint [0]{\href }%
\providecommand \doibase [0]{https://doi.org/}%
\providecommand \selectlanguage [0]{\@gobble}%
\providecommand \bibinfo  [0]{\@secondoftwo}%
\providecommand \bibfield  [0]{\@secondoftwo}%
\providecommand \translation [1]{[#1]}%
\providecommand \BibitemOpen [0]{}%
\providecommand \bibitemStop [0]{}%
\providecommand \bibitemNoStop [0]{.\EOS\space}%
\providecommand \EOS [0]{\spacefactor3000\relax}%
\providecommand \BibitemShut  [1]{\csname bibitem#1\endcsname}%
\let\auto@bib@innerbib\@empty
%</preamble>
\bibitem [{\citenamefont {Mahan}(2000)}]{mahan_many-particle_2000}%
  \BibitemOpen
  \bibfield  {author} {\bibinfo {author} {\bibfnamefont {G.}~\bibnamefont
  {Mahan}},\ }\href@noop {} {\emph {\bibinfo {title} {Many-particle
  physics}}},\ \bibinfo {edition} {3rd}\ ed.\ (\bibinfo  {publisher} {Kluwer
  Academic/Plenum Publishers},\ \bibinfo {address} {New York},\ \bibinfo {year}
  {2000})\BibitemShut {NoStop}%
\bibitem [{\citenamefont {van
  Leeuwen}(2004)}]{van_leeuwen_first-principles_2004}%
  \BibitemOpen
  \bibfield  {author} {\bibinfo {author} {\bibfnamefont {R.}~\bibnamefont {van
  Leeuwen}},\ }\bibfield  {title} {\bibinfo {title} {First-principles approach
  to the electron-phonon interaction},\ }\href
  {https://doi.org/10.1103/PhysRevB.69.115110} {\bibfield  {journal} {\bibinfo
  {journal} {Phys. Rev. B}\ }\textbf {\bibinfo {volume} {69}},\ \bibinfo
  {pages} {115110} (\bibinfo {year} {2004})}\BibitemShut {NoStop}%
\bibitem [{\citenamefont {Giustino}(2017)}]{giustino_electron-phonon_2017}%
  \BibitemOpen
  \bibfield  {author} {\bibinfo {author} {\bibfnamefont {F.}~\bibnamefont
  {Giustino}},\ }\bibfield  {title} {\bibinfo {title} {Electron-phonon
  interactions from first principles},\ }\href
  {https://doi.org/10.1103/RevModPhys.89.015003} {\bibfield  {journal}
  {\bibinfo  {journal} {Rev. Mod. Phys.}\ }\textbf {\bibinfo {volume} {89}},\
  \bibinfo {pages} {015003} (\bibinfo {year} {2017})}\BibitemShut {NoStop}%
\bibitem [{\citenamefont {Langreth}(1970)}]{langreth_singularities_1970}%
  \BibitemOpen
  \bibfield  {author} {\bibinfo {author} {\bibfnamefont {D.~C.}\ \bibnamefont
  {Langreth}},\ }\bibfield  {title} {\bibinfo {title} {Singularities in the
  {X}-{Ray} {Spectra} of {Metals}},\ }\href
  {https://doi.org/10.1103/PhysRevB.1.471} {\bibfield  {journal} {\bibinfo
  {journal} {Phys. Rev. B}\ }\textbf {\bibinfo {volume} {1}},\ \bibinfo {pages}
  {471} (\bibinfo {year} {1970})}\BibitemShut {NoStop}%
\bibitem [{\citenamefont {Sch\"{u}ler}\ \emph {et~al.}(2016)\citenamefont
  {Sch\"{u}ler}, \citenamefont {Berakdar},\ and\ \citenamefont
  {Pavlyukh}}]{schuler_time-dependent_2016}%
  \BibitemOpen
  \bibfield  {author} {\bibinfo {author} {\bibfnamefont {M.}~\bibnamefont
  {Sch\"{u}ler}}, \bibinfo {author} {\bibfnamefont {J.}~\bibnamefont
  {Berakdar}},\ and\ \bibinfo {author} {\bibfnamefont {Y.}~\bibnamefont
  {Pavlyukh}},\ }\bibfield  {title} {\bibinfo {title} {Time-dependent many-body
  treatment of electron-boson dynamics: {Application} to plasmon-accompanied
  photoemission},\ }\href {https://doi.org/10.1103/PhysRevB.93.054303}
  {\bibfield  {journal} {\bibinfo  {journal} {Phys. Rev. B}\ }\textbf {\bibinfo
  {volume} {93}},\ \bibinfo {pages} {054303} (\bibinfo {year}
  {2016})}\BibitemShut {NoStop}%
\bibitem [{\citenamefont {Pavlyukh}(2017)}]{pavlyukh_pade_2017}%
  \BibitemOpen
  \bibfield  {author} {\bibinfo {author} {\bibfnamefont {Y.}~\bibnamefont
  {Pavlyukh}},\ }\bibfield  {title} {\bibinfo {title} {Pad\'{e} resummation of
  many-body perturbation theories},\ }\href
  {https://doi.org/10.1038/s41598-017-00355-w} {\bibfield  {journal} {\bibinfo
  {journal} {Sci. Rep.}\ }\textbf {\bibinfo {volume} {7}},\ \bibinfo {pages}
  {504} (\bibinfo {year} {2017})}\BibitemShut {NoStop}%
\bibitem [{\citenamefont {S\"{a}kkinen}\ \emph {et~al.}(2015)\citenamefont
  {S\"{a}kkinen}, \citenamefont {Peng}, \citenamefont {Appel},\ and\
  \citenamefont {van Leeuwen}}]{sakkinen_many-body_2015}%
  \BibitemOpen
  \bibfield  {author} {\bibinfo {author} {\bibfnamefont {N.}~\bibnamefont
  {S\"{a}kkinen}}, \bibinfo {author} {\bibfnamefont {Y.}~\bibnamefont {Peng}},
  \bibinfo {author} {\bibfnamefont {H.}~\bibnamefont {Appel}},\ and\ \bibinfo
  {author} {\bibfnamefont {R.}~\bibnamefont {van Leeuwen}},\ }\bibfield
  {title} {\bibinfo {title} {Many-body {Green}'s function theory for
  electron-phonon interactions: {The} {Kadanoff}-{Baym} approach to spectral
  properties of the {Holstein} dimer},\ }\href
  {https://doi.org/10.1063/1.4936143} {\bibfield  {journal} {\bibinfo
  {journal} {J. Chem. Phys.}\ }\textbf {\bibinfo {volume} {143}},\ \bibinfo
  {pages} {234102} (\bibinfo {year} {2015})}\BibitemShut {NoStop}%
\bibitem [{\citenamefont {Tuovinen}\ \emph {et~al.}(2016)\citenamefont
  {Tuovinen}, \citenamefont {S\"{a}kkinen}, \citenamefont {Karlsson},
  \citenamefont {Stefanucci},\ and\ \citenamefont {van
  Leeuwen}}]{tuovinen_phononic_2016}%
  \BibitemOpen
  \bibfield  {author} {\bibinfo {author} {\bibfnamefont {R.}~\bibnamefont
  {Tuovinen}}, \bibinfo {author} {\bibfnamefont {N.}~\bibnamefont
  {S\"{a}kkinen}}, \bibinfo {author} {\bibfnamefont {D.}~\bibnamefont
  {Karlsson}}, \bibinfo {author} {\bibfnamefont {G.}~\bibnamefont
  {Stefanucci}},\ and\ \bibinfo {author} {\bibfnamefont {R.}~\bibnamefont {van
  Leeuwen}},\ }\bibfield  {title} {\bibinfo {title} {Phononic heat transport in
  the transient regime: {An} analytic solution},\ }\href
  {https://doi.org/10.1103/PhysRevB.93.214301} {\bibfield  {journal} {\bibinfo
  {journal} {Phys. Rev. B}\ }\textbf {\bibinfo {volume} {93}},\ \bibinfo
  {pages} {214301} (\bibinfo {year} {2016})}\BibitemShut {NoStop}%
\bibitem [{\citenamefont {Galperin}\ \emph {et~al.}(2008)\citenamefont
  {Galperin}, \citenamefont {Nitzan},\ and\ \citenamefont
  {Ratner}}]{galperin_non-linear_2008}%
  \BibitemOpen
  \bibfield  {author} {\bibinfo {author} {\bibfnamefont {M.}~\bibnamefont
  {Galperin}}, \bibinfo {author} {\bibfnamefont {A.}~\bibnamefont {Nitzan}},\
  and\ \bibinfo {author} {\bibfnamefont {M.~A.}\ \bibnamefont {Ratner}},\
  }\bibfield  {title} {\bibinfo {title} {The non-linear response of molecular
  junctions: the polaron model revisited},\ }\href
  {https://doi.org/10.1088/0953-8984/20/37/374107} {\bibfield  {journal}
  {\bibinfo  {journal} {J. Phys. Condens. Matter}\ }\textbf {\bibinfo {volume}
  {20}},\ \bibinfo {pages} {374107} (\bibinfo {year} {2008})}\BibitemShut
  {NoStop}%
\bibitem [{\citenamefont {Leturcq}\ \emph {et~al.}(2009)\citenamefont
  {Leturcq}, \citenamefont {Stampfer}, \citenamefont {Inderbitzin},
  \citenamefont {Durrer}, \citenamefont {Hierold}, \citenamefont {Mariani},
  \citenamefont {Schultz}, \citenamefont {von Oppen},\ and\ \citenamefont
  {Ensslin}}]{leturcq_franckcondon_2009}%
  \BibitemOpen
  \bibfield  {author} {\bibinfo {author} {\bibfnamefont {R.}~\bibnamefont
  {Leturcq}}, \bibinfo {author} {\bibfnamefont {C.}~\bibnamefont {Stampfer}},
  \bibinfo {author} {\bibfnamefont {K.}~\bibnamefont {Inderbitzin}}, \bibinfo
  {author} {\bibfnamefont {L.}~\bibnamefont {Durrer}}, \bibinfo {author}
  {\bibfnamefont {C.}~\bibnamefont {Hierold}}, \bibinfo {author} {\bibfnamefont
  {E.}~\bibnamefont {Mariani}}, \bibinfo {author} {\bibfnamefont {M.~G.}\
  \bibnamefont {Schultz}}, \bibinfo {author} {\bibfnamefont {F.}~\bibnamefont
  {von Oppen}},\ and\ \bibinfo {author} {\bibfnamefont {K.}~\bibnamefont
  {Ensslin}},\ }\bibfield  {title} {\bibinfo {title} {Franck-{Condon} blockade
  in suspended carbon nanotube quantum dots},\ }\href
  {https://doi.org/10.1038/nphys1234} {\bibfield  {journal} {\bibinfo
  {journal} {Nature Phys.}\ }\textbf {\bibinfo {volume} {5}},\ \bibinfo {pages}
  {327} (\bibinfo {year} {2009})}\BibitemShut {NoStop}%
\bibitem [{\citenamefont {White}\ and\ \citenamefont
  {Galperin}(2012)}]{white_inelastic_2012}%
  \BibitemOpen
  \bibfield  {author} {\bibinfo {author} {\bibfnamefont {A.~J.}\ \bibnamefont
  {White}}\ and\ \bibinfo {author} {\bibfnamefont {M.}~\bibnamefont
  {Galperin}},\ }\bibfield  {title} {\bibinfo {title} {Inelastic transport: a
  pseudoparticle approach},\ }\href {https://doi.org/10.1039/C2CP41017F}
  {\bibfield  {journal} {\bibinfo  {journal} {Phys. Chem. Chem. Phys.}\
  }\textbf {\bibinfo {volume} {14}},\ \bibinfo {pages} {13809} (\bibinfo {year}
  {2012})}\BibitemShut {NoStop}%
\bibitem [{\citenamefont {Perfetto}\ and\ \citenamefont
  {Stefanucci}(2013)}]{perfetto_image_2013}%
  \BibitemOpen
  \bibfield  {author} {\bibinfo {author} {\bibfnamefont {E.}~\bibnamefont
  {Perfetto}}\ and\ \bibinfo {author} {\bibfnamefont {G.}~\bibnamefont
  {Stefanucci}},\ }\bibfield  {title} {\bibinfo {title} {Image charge effects
  in the nonequilibrium {Anderson}-{Holstein} model},\ }\href
  {https://doi.org/10.1103/PhysRevB.88.245437} {\bibfield  {journal} {\bibinfo
  {journal} {Phys. Rev. B}\ }\textbf {\bibinfo {volume} {88}},\ \bibinfo
  {pages} {245437} (\bibinfo {year} {2013})}\BibitemShut {NoStop}%
\bibitem [{\citenamefont {Wilner}\ \emph {et~al.}(2014)\citenamefont {Wilner},
  \citenamefont {Wang}, \citenamefont {Thoss},\ and\ \citenamefont
  {Rabani}}]{wilner_nonequilibrium_2014}%
  \BibitemOpen
  \bibfield  {author} {\bibinfo {author} {\bibfnamefont {E.~Y.}\ \bibnamefont
  {Wilner}}, \bibinfo {author} {\bibfnamefont {H.}~\bibnamefont {Wang}},
  \bibinfo {author} {\bibfnamefont {M.}~\bibnamefont {Thoss}},\ and\ \bibinfo
  {author} {\bibfnamefont {E.}~\bibnamefont {Rabani}},\ }\bibfield  {title}
  {\bibinfo {title} {Nonequilibrium quantum systems with electron-phonon
  interactions: {Transient} dynamics and approach to steady state},\ }\href
  {https://doi.org/10.1103/PhysRevB.89.205129} {\bibfield  {journal} {\bibinfo
  {journal} {Phys. Rev. B}\ }\textbf {\bibinfo {volume} {89}},\ \bibinfo
  {pages} {205129} (\bibinfo {year} {2014})}\BibitemShut {NoStop}%
\bibitem [{\citenamefont {M\"{u}hlbacher}\ and\ \citenamefont
  {Rabani}(2008)}]{muhlbacher_real-time_2008}%
  \BibitemOpen
  \bibfield  {author} {\bibinfo {author} {\bibfnamefont {L.}~\bibnamefont
  {M\"{u}hlbacher}}\ and\ \bibinfo {author} {\bibfnamefont {E.}~\bibnamefont
  {Rabani}},\ }\bibfield  {title} {\bibinfo {title} {Real-{Time} {Path}
  {Integral} {Approach} to {Nonequilibrium} {Many}-{Body} {Quantum}
  {Systems}},\ }\href {https://doi.org/10.1103/PhysRevLett.100.176403}
  {\bibfield  {journal} {\bibinfo  {journal} {Phys. Rev. Lett.}\ }\textbf
  {\bibinfo {volume} {100}},\ \bibinfo {pages} {176403} (\bibinfo {year}
  {2008})}\BibitemShut {NoStop}%
\bibitem [{\citenamefont {Galperin}(2017)}]{galperin_photonics_2017}%
  \BibitemOpen
  \bibfield  {author} {\bibinfo {author} {\bibfnamefont {M.}~\bibnamefont
  {Galperin}},\ }\bibfield  {title} {\bibinfo {title} {Photonics and
  spectroscopy in nanojunctions: a theoretical insight},\ }\href
  {https://doi.org/10.1039/C7CS00067G} {\bibfield  {journal} {\bibinfo
  {journal} {Chem. Soc. Rev.}\ }\textbf {\bibinfo {volume} {46}},\ \bibinfo
  {pages} {4000} (\bibinfo {year} {2017})}\BibitemShut {NoStop}%
\bibitem [{\citenamefont {Chen}\ \emph {et~al.}(2015)\citenamefont {Chen},
  \citenamefont {Zhao},\ and\ \citenamefont {Tanimura}}]{chen_dynamics_2015}%
  \BibitemOpen
  \bibfield  {author} {\bibinfo {author} {\bibfnamefont {L.}~\bibnamefont
  {Chen}}, \bibinfo {author} {\bibfnamefont {Y.}~\bibnamefont {Zhao}},\ and\
  \bibinfo {author} {\bibfnamefont {Y.}~\bibnamefont {Tanimura}},\ }\bibfield
  {title} {\bibinfo {title} {Dynamics of a {One}-{Dimensional} {Holstein}
  {Polaron} with the {Hierarchical} {Equations} of {Motion} {Approach}},\
  }\href {https://doi.org/10.1021/acs.jpclett.5b01368} {\bibfield  {journal}
  {\bibinfo  {journal} {J. Phys. Chem. Lett.}\ }\textbf {\bibinfo {volume}
  {6}},\ \bibinfo {pages} {3110} (\bibinfo {year} {2015})}\BibitemShut
  {NoStop}%
\bibitem [{\citenamefont {Tanimura}(2020)}]{tanimura_numerically_2020}%
  \BibitemOpen
  \bibfield  {author} {\bibinfo {author} {\bibfnamefont {Y.}~\bibnamefont
  {Tanimura}},\ }\bibfield  {title} {\bibinfo {title} {Numerically “exact”
  approach to open quantum dynamics: {The} hierarchical equations of motion
  ({HEOM})},\ }\href {https://doi.org/10.1063/5.0011599} {\bibfield  {journal}
  {\bibinfo  {journal} {J. Chem. Phys.}\ }\textbf {\bibinfo {volume} {153}},\
  \bibinfo {pages} {020901} (\bibinfo {year} {2020})}\BibitemShut {NoStop}%
\bibitem [{\citenamefont {Werdehausen}\ \emph {et~al.}(2018)\citenamefont
  {Werdehausen}, \citenamefont {Takayama}, \citenamefont {H\"{o}ppner},
  \citenamefont {Albrecht}, \citenamefont {Rost}, \citenamefont {Lu},
  \citenamefont {Manske}, \citenamefont {Takagi},\ and\ \citenamefont
  {Kaiser}}]{werdehausen_coherent_2018}%
  \BibitemOpen
  \bibfield  {author} {\bibinfo {author} {\bibfnamefont {D.}~\bibnamefont
  {Werdehausen}}, \bibinfo {author} {\bibfnamefont {T.}~\bibnamefont
  {Takayama}}, \bibinfo {author} {\bibfnamefont {M.}~\bibnamefont
  {H\"{o}ppner}}, \bibinfo {author} {\bibfnamefont {G.}~\bibnamefont
  {Albrecht}}, \bibinfo {author} {\bibfnamefont {A.~W.}\ \bibnamefont {Rost}},
  \bibinfo {author} {\bibfnamefont {Y.}~\bibnamefont {Lu}}, \bibinfo {author}
  {\bibfnamefont {D.}~\bibnamefont {Manske}}, \bibinfo {author} {\bibfnamefont
  {H.}~\bibnamefont {Takagi}},\ and\ \bibinfo {author} {\bibfnamefont
  {S.}~\bibnamefont {Kaiser}},\ }\bibfield  {title} {\bibinfo {title} {Coherent
  order parameter oscillations in the ground state of the excitonic insulator
  {Ta}$_{\textrm{2}}${NiSe}$_{\textrm{5}}$},\ }\href
  {https://doi.org/10.1126/sciadv.aap8652} {\bibfield  {journal} {\bibinfo
  {journal} {Science Advances}\ }\textbf {\bibinfo {volume} {4}},\ \bibinfo
  {pages} {eaap8652} (\bibinfo {year} {2018})}\BibitemShut {NoStop}%
\bibitem [{\citenamefont {Kemper}\ \emph {et~al.}(2015)\citenamefont {Kemper},
  \citenamefont {Sentef}, \citenamefont {Moritz}, \citenamefont {Freericks},\
  and\ \citenamefont {Devereaux}}]{kemper_direct_2015}%
  \BibitemOpen
  \bibfield  {author} {\bibinfo {author} {\bibfnamefont {A.~F.}\ \bibnamefont
  {Kemper}}, \bibinfo {author} {\bibfnamefont {M.~A.}\ \bibnamefont {Sentef}},
  \bibinfo {author} {\bibfnamefont {B.}~\bibnamefont {Moritz}}, \bibinfo
  {author} {\bibfnamefont {J.~K.}\ \bibnamefont {Freericks}},\ and\ \bibinfo
  {author} {\bibfnamefont {T.~P.}\ \bibnamefont {Devereaux}},\ }\bibfield
  {title} {\bibinfo {title} {Direct observation of {Higgs} mode oscillations in
  the pump-probe photoemission spectra of electron-phonon mediated
  superconductors},\ }\href {https://doi.org/10.1103/PhysRevB.92.224517}
  {\bibfield  {journal} {\bibinfo  {journal} {Phys. Rev. B}\ }\textbf {\bibinfo
  {volume} {92}},\ \bibinfo {pages} {224517} (\bibinfo {year}
  {2015})}\BibitemShut {NoStop}%
\bibitem [{\citenamefont {Stolpp}\ \emph {et~al.}(2020)\citenamefont {Stolpp},
  \citenamefont {Herbrych}, \citenamefont {Dorfner}, \citenamefont {Dagotto},\
  and\ \citenamefont {Heidrich-Meisner}}]{stolpp_charge-density-wave_2020}%
  \BibitemOpen
  \bibfield  {author} {\bibinfo {author} {\bibfnamefont {J.}~\bibnamefont
  {Stolpp}}, \bibinfo {author} {\bibfnamefont {J.}~\bibnamefont {Herbrych}},
  \bibinfo {author} {\bibfnamefont {F.}~\bibnamefont {Dorfner}}, \bibinfo
  {author} {\bibfnamefont {E.}~\bibnamefont {Dagotto}},\ and\ \bibinfo {author}
  {\bibfnamefont {F.}~\bibnamefont {Heidrich-Meisner}},\ }\bibfield  {title}
  {\bibinfo {title} {Charge-density-wave melting in the one-dimensional
  {Holstein} model},\ }\href {https://doi.org/10.1103/PhysRevB.101.035134}
  {\bibfield  {journal} {\bibinfo  {journal} {Phys. Rev. B}\ }\textbf {\bibinfo
  {volume} {101}},\ \bibinfo {pages} {035134} (\bibinfo {year}
  {2020})}\BibitemShut {NoStop}%
\bibitem [{\citenamefont {Devreese}\ and\ \citenamefont
  {Alexandrov}(2009)}]{devreese_frohlich_2009}%
  \BibitemOpen
  \bibfield  {author} {\bibinfo {author} {\bibfnamefont {J.~T.}\ \bibnamefont
  {Devreese}}\ and\ \bibinfo {author} {\bibfnamefont {A.~S.}\ \bibnamefont
  {Alexandrov}},\ }\bibfield  {title} {\bibinfo {title} {Fr\"{o}hlich polaron
  and bipolaron: recent developments},\ }\href
  {https://doi.org/10.1088/0034-4885/72/6/066501} {\bibfield  {journal}
  {\bibinfo  {journal} {Rep. Prog. Phys.}\ }\textbf {\bibinfo {volume} {72}},\
  \bibinfo {pages} {066501} (\bibinfo {year} {2009})}\BibitemShut {NoStop}%
\bibitem [{\citenamefont {Leong}\ and\ \citenamefont
  {Vittal}(2011)}]{leong_one-dimensional_2011}%
  \BibitemOpen
  \bibfield  {author} {\bibinfo {author} {\bibfnamefont {W.~L.}\ \bibnamefont
  {Leong}}\ and\ \bibinfo {author} {\bibfnamefont {J.~J.}\ \bibnamefont
  {Vittal}},\ }\bibfield  {title} {\bibinfo {title} {One-{Dimensional}
  {Coordination} {Polymers}: {Complexity} and {Diversity} in {Structures},
  {Properties}, and {Applications}},\ }\href
  {https://doi.org/10.1021/cr100160e} {\bibfield  {journal} {\bibinfo
  {journal} {Chem. Rev.}\ }\textbf {\bibinfo {volume} {111}},\ \bibinfo {pages}
  {688} (\bibinfo {year} {2011})}\BibitemShut {NoStop}%
\bibitem [{\citenamefont {Dexheimer}\ \emph {et~al.}(2000)\citenamefont
  {Dexheimer}, \citenamefont {Van~Pelt}, \citenamefont {Brozik},\ and\
  \citenamefont {Swanson}}]{dexheimer_femtosecond_2000}%
  \BibitemOpen
  \bibfield  {author} {\bibinfo {author} {\bibfnamefont {S.~L.}\ \bibnamefont
  {Dexheimer}}, \bibinfo {author} {\bibfnamefont {A.~D.}\ \bibnamefont
  {Van~Pelt}}, \bibinfo {author} {\bibfnamefont {J.~A.}\ \bibnamefont
  {Brozik}},\ and\ \bibinfo {author} {\bibfnamefont {B.~I.}\ \bibnamefont
  {Swanson}},\ }\bibfield  {title} {\bibinfo {title} {Femtosecond {Vibrational}
  {Dynamics} of {Self}-{Trapping} in a {Quasi}-{One}-{Dimensional} {System}},\
  }\href {https://doi.org/10.1103/PhysRevLett.84.4425} {\bibfield  {journal}
  {\bibinfo  {journal} {Phys. Rev. Lett.}\ }\textbf {\bibinfo {volume} {84}},\
  \bibinfo {pages} {4425} (\bibinfo {year} {2000})}\BibitemShut {NoStop}%
\bibitem [{\citenamefont {Sugita}\ \emph {et~al.}(2001)\citenamefont {Sugita},
  \citenamefont {Saito}, \citenamefont {Kano}, \citenamefont {Yamashita},\ and\
  \citenamefont {Kobayashi}}]{sugita_wave_2001}%
  \BibitemOpen
  \bibfield  {author} {\bibinfo {author} {\bibfnamefont {A.}~\bibnamefont
  {Sugita}}, \bibinfo {author} {\bibfnamefont {T.}~\bibnamefont {Saito}},
  \bibinfo {author} {\bibfnamefont {H.}~\bibnamefont {Kano}}, \bibinfo {author}
  {\bibfnamefont {M.}~\bibnamefont {Yamashita}},\ and\ \bibinfo {author}
  {\bibfnamefont {T.}~\bibnamefont {Kobayashi}},\ }\bibfield  {title} {\bibinfo
  {title} {Wave {Packet} {Dynamics} in a {Quasi}-{One}-{Dimensional}
  {Metal}-{Halogen} {Complex} {Studied} by {Ultrafast} {Time}-{Resolved}
  {Spectroscopy}},\ }\href {https://doi.org/10.1103/PhysRevLett.86.2158}
  {\bibfield  {journal} {\bibinfo  {journal} {Phys. Rev. Lett.}\ }\textbf
  {\bibinfo {volume} {86}},\ \bibinfo {pages} {2158} (\bibinfo {year}
  {2001})}\BibitemShut {NoStop}%
\bibitem [{\citenamefont {Ku}\ and\ \citenamefont
  {Trugman}(2007)}]{ku_quantum_2007}%
  \BibitemOpen
  \bibfield  {author} {\bibinfo {author} {\bibfnamefont {L.-C.}\ \bibnamefont
  {Ku}}\ and\ \bibinfo {author} {\bibfnamefont {S.~A.}\ \bibnamefont
  {Trugman}},\ }\bibfield  {title} {\bibinfo {title} {Quantum dynamics of
  polaron formation},\ }\href {https://doi.org/10.1103/PhysRevB.75.014307}
  {\bibfield  {journal} {\bibinfo  {journal} {Phys. Rev. B}\ }\textbf {\bibinfo
  {volume} {75}},\ \bibinfo {pages} {014307} (\bibinfo {year}
  {2007})}\BibitemShut {NoStop}%
\bibitem [{\citenamefont {Gole\v{z}}\ \emph
  {et~al.}(2012{\natexlab{a}})\citenamefont {Gole\v{z}}, \citenamefont
  {Bon\v{c}a}, \citenamefont {Vidmar},\ and\ \citenamefont
  {Trugman}}]{golez_relaxation_2012}%
  \BibitemOpen
  \bibfield  {author} {\bibinfo {author} {\bibfnamefont {D.}~\bibnamefont
  {Gole\v{z}}}, \bibinfo {author} {\bibfnamefont {J.}~\bibnamefont
  {Bon\v{c}a}}, \bibinfo {author} {\bibfnamefont {L.}~\bibnamefont {Vidmar}},\
  and\ \bibinfo {author} {\bibfnamefont {S.~A.}\ \bibnamefont {Trugman}},\
  }\bibfield  {title} {\bibinfo {title} {Relaxation {Dynamics} of the
  {Holstein} {Polaron}},\ }\href
  {https://doi.org/10.1103/PhysRevLett.109.236402} {\bibfield  {journal}
  {\bibinfo  {journal} {Phys. Rev. Lett.}\ }\textbf {\bibinfo {volume} {109}},\
  \bibinfo {pages} {236402} (\bibinfo {year} {2012}{\natexlab{a}})}\BibitemShut
  {NoStop}%
\bibitem [{\citenamefont {Gole\v{z}}\ \emph
  {et~al.}(2012{\natexlab{b}})\citenamefont {Gole\v{z}}, \citenamefont
  {Bon\v{c}a},\ and\ \citenamefont {Vidmar}}]{golez_dissociation_2012}%
  \BibitemOpen
  \bibfield  {author} {\bibinfo {author} {\bibfnamefont {D.}~\bibnamefont
  {Gole\v{z}}}, \bibinfo {author} {\bibfnamefont {J.}~\bibnamefont
  {Bon\v{c}a}},\ and\ \bibinfo {author} {\bibfnamefont {L.}~\bibnamefont
  {Vidmar}},\ }\bibfield  {title} {\bibinfo {title} {Dissociation of a
  {Hubbard}-{Holstein} bipolaron driven away from equilibrium by a constant
  electric field},\ }\href {https://doi.org/10.1103/PhysRevB.85.144304}
  {\bibfield  {journal} {\bibinfo  {journal} {Phys. Rev. B}\ }\textbf {\bibinfo
  {volume} {85}},\ \bibinfo {pages} {144304} (\bibinfo {year}
  {2012}{\natexlab{b}})}\BibitemShut {NoStop}%
\bibitem [{\citenamefont {Dorfner}\ \emph {et~al.}(2015)\citenamefont
  {Dorfner}, \citenamefont {Vidmar}, \citenamefont {Brockt}, \citenamefont
  {Jeckelmann},\ and\ \citenamefont
  {Heidrich-Meisner}}]{dorfner_real-time_2015}%
  \BibitemOpen
  \bibfield  {author} {\bibinfo {author} {\bibfnamefont {F.}~\bibnamefont
  {Dorfner}}, \bibinfo {author} {\bibfnamefont {L.}~\bibnamefont {Vidmar}},
  \bibinfo {author} {\bibfnamefont {C.}~\bibnamefont {Brockt}}, \bibinfo
  {author} {\bibfnamefont {E.}~\bibnamefont {Jeckelmann}},\ and\ \bibinfo
  {author} {\bibfnamefont {F.}~\bibnamefont {Heidrich-Meisner}},\ }\bibfield
  {title} {\bibinfo {title} {Real-time decay of a highly excited charge carrier
  in the one-dimensional {Holstein} model},\ }\href
  {https://doi.org/10.1103/PhysRevB.91.104302} {\bibfield  {journal} {\bibinfo
  {journal} {Phys. Rev. B}\ }\textbf {\bibinfo {volume} {91}},\ \bibinfo
  {pages} {104302} (\bibinfo {year} {2015})}\BibitemShut {NoStop}%
\bibitem [{\citenamefont {Rausch}\ and\ \citenamefont
  {Potthoff}(2017)}]{rausch_filling-dependent_2017}%
  \BibitemOpen
  \bibfield  {author} {\bibinfo {author} {\bibfnamefont {R.}~\bibnamefont
  {Rausch}}\ and\ \bibinfo {author} {\bibfnamefont {M.}~\bibnamefont
  {Potthoff}},\ }\bibfield  {title} {\bibinfo {title} {Filling-dependent
  doublon dynamics in the one-dimensional {Hubbard} model},\ }\href
  {https://doi.org/10.1103/PhysRevB.95.045152} {\bibfield  {journal} {\bibinfo
  {journal} {Phys. Rev. B}\ }\textbf {\bibinfo {volume} {95}},\ \bibinfo
  {pages} {045152} (\bibinfo {year} {2017})}\BibitemShut {NoStop}%
\bibitem [{\citenamefont {Kloss}\ \emph {et~al.}(2019)\citenamefont {Kloss},
  \citenamefont {Reichman},\ and\ \citenamefont
  {Tempelaar}}]{kloss_multiset_2019}%
  \BibitemOpen
  \bibfield  {author} {\bibinfo {author} {\bibfnamefont {B.}~\bibnamefont
  {Kloss}}, \bibinfo {author} {\bibfnamefont {D.~R.}\ \bibnamefont
  {Reichman}},\ and\ \bibinfo {author} {\bibfnamefont {R.}~\bibnamefont
  {Tempelaar}},\ }\bibfield  {title} {\bibinfo {title} {Multiset {Matrix}
  {Product} {State} {Calculations} {Reveal} {Mobile} {Franck}-{Condon}
  {Excitations} {Under} {Strong} {Holstein}-{Type} {Coupling}},\ }\href
  {https://doi.org/10.1103/PhysRevLett.123.126601} {\bibfield  {journal}
  {\bibinfo  {journal} {Phys. Rev. Lett.}\ }\textbf {\bibinfo {volume} {123}},\
  \bibinfo {pages} {126601} (\bibinfo {year} {2019})}\BibitemShut {NoStop}%
\bibitem [{\citenamefont {Frahm}\ and\ \citenamefont
  {Pfannkuche}(2019)}]{frahm_ultrafast_2019}%
  \BibitemOpen
  \bibfield  {author} {\bibinfo {author} {\bibfnamefont {L.-H.}\ \bibnamefont
  {Frahm}}\ and\ \bibinfo {author} {\bibfnamefont {D.}~\bibnamefont
  {Pfannkuche}},\ }\bibfield  {title} {\bibinfo {title} {Ultrafast ab {Initio}
  {Quantum} {Chemistry} {Using} {Matrix} {Product} {States}},\ }\href
  {https://doi.org/10.1021/acs.jctc.8b01291} {\bibfield  {journal} {\bibinfo
  {journal} {J. Chem. Theory Comput.}\ }\textbf {\bibinfo {volume} {15}},\
  \bibinfo {pages} {2154} (\bibinfo {year} {2019})}\BibitemShut {NoStop}%
\bibitem [{\citenamefont {Ponseca}\ \emph {et~al.}(2017)\citenamefont
  {Ponseca}, \citenamefont {Ch\'{a}bera}, \citenamefont {Uhlig}, \citenamefont
  {Persson},\ and\ \citenamefont {Sundstr\"{o}m}}]{ponseca_ultrafast_2017}%
  \BibitemOpen
  \bibfield  {author} {\bibinfo {author} {\bibfnamefont {C.~S.}\ \bibnamefont
  {Ponseca}}, \bibinfo {author} {\bibfnamefont {P.}~\bibnamefont
  {Ch\'{a}bera}}, \bibinfo {author} {\bibfnamefont {J.}~\bibnamefont {Uhlig}},
  \bibinfo {author} {\bibfnamefont {P.}~\bibnamefont {Persson}},\ and\ \bibinfo
  {author} {\bibfnamefont {V.}~\bibnamefont {Sundstr\"{o}m}},\ }\bibfield
  {title} {\bibinfo {title} {Ultrafast {Electron} {Dynamics} in {Solar}
  {Energy} {Conversion}},\ }\href {https://doi.org/10.1021/acs.chemrev.6b00807}
  {\bibfield  {journal} {\bibinfo  {journal} {Chem. Rev.}\ }\textbf {\bibinfo
  {volume} {117}},\ \bibinfo {pages} {10940} (\bibinfo {year}
  {2017})}\BibitemShut {NoStop}%
\bibitem [{\citenamefont {Pastor}\ \emph {et~al.}(2019)\citenamefont {Pastor},
  \citenamefont {Park}, \citenamefont {Steier}, \citenamefont {Kim},
  \citenamefont {Gr\"{a}tzel}, \citenamefont {Durrant}, \citenamefont {Walsh},\
  and\ \citenamefont {Bakulin}}]{pastor_situ_2019}%
  \BibitemOpen
  \bibfield  {author} {\bibinfo {author} {\bibfnamefont {E.}~\bibnamefont
  {Pastor}}, \bibinfo {author} {\bibfnamefont {J.-S.}\ \bibnamefont {Park}},
  \bibinfo {author} {\bibfnamefont {L.}~\bibnamefont {Steier}}, \bibinfo
  {author} {\bibfnamefont {S.}~\bibnamefont {Kim}}, \bibinfo {author}
  {\bibfnamefont {M.}~\bibnamefont {Gr\"{a}tzel}}, \bibinfo {author}
  {\bibfnamefont {J.~R.}\ \bibnamefont {Durrant}}, \bibinfo {author}
  {\bibfnamefont {A.}~\bibnamefont {Walsh}},\ and\ \bibinfo {author}
  {\bibfnamefont {A.~A.}\ \bibnamefont {Bakulin}},\ }\bibfield  {title}
  {\bibinfo {title} {In situ observation of picosecond polaron
  self-localisation in $\alpha$-{Fe}$_{\textrm{2}}${O}$_{\textrm{3}}$
  photoelectrochemical cells},\ }\href
  {https://doi.org/10.1038/s41467-019-11767-9} {\bibfield  {journal} {\bibinfo
  {journal} {Nat. Commun.}\ }\textbf {\bibinfo {volume} {10}},\ \bibinfo
  {pages} {3962} (\bibinfo {year} {2019})}\BibitemShut {NoStop}%
\bibitem [{\citenamefont {Karlsson}\ \emph {et~al.}(2021)\citenamefont
  {Karlsson}, \citenamefont {van Leeuwen}, \citenamefont {Pavlyukh},
  \citenamefont {Perfetto},\ and\ \citenamefont
  {Stefanucci}}]{karlsson_fast_2021}%
  \BibitemOpen
  \bibfield  {author} {\bibinfo {author} {\bibfnamefont {D.}~\bibnamefont
  {Karlsson}}, \bibinfo {author} {\bibfnamefont {R.}~\bibnamefont {van
  Leeuwen}}, \bibinfo {author} {\bibfnamefont {Y.}~\bibnamefont {Pavlyukh}},
  \bibinfo {author} {\bibfnamefont {E.}~\bibnamefont {Perfetto}},\ and\
  \bibinfo {author} {\bibfnamefont {G.}~\bibnamefont {Stefanucci}},\ }\bibfield
   {title} {\bibinfo {title} {Fast {Green}'s {Function} {Method} for
  {Ultrafast} {Electron}-{Boson} {Dynamics}},\ }\href
  {https://doi.org/10.1103/PhysRevLett.127.036402} {\bibfield  {journal}
  {\bibinfo  {journal} {Phys. Rev. Lett.}\ }\textbf {\bibinfo {volume} {127}},\
  \bibinfo {pages} {036402} (\bibinfo {year} {2021})}\BibitemShut {NoStop}%
\bibitem [{\citenamefont {Perfetto}\ \emph {et~al.}(2021)\citenamefont
  {Perfetto}, \citenamefont {Pavlyukh},\ and\ \citenamefont
  {Stefanucci}}]{perfetto_tdgw_2021}%
  \BibitemOpen
  \bibfield  {author} {\bibinfo {author} {\bibfnamefont {E.}~\bibnamefont
  {Perfetto}}, \bibinfo {author} {\bibfnamefont {Y.}~\bibnamefont {Pavlyukh}},\
  and\ \bibinfo {author} {\bibfnamefont {G.}~\bibnamefont {Stefanucci}},\
  }\bibfield  {title} {\bibinfo {title} {Real-time {$GW$}: \emph{ab initio}
  description of the ultrafast carrier and exciton dynamics in two-dimensional
  systems},\ }\href {http://arxiv.org/abs/2109.15209} {\bibfield  {journal}
  {\bibinfo  {journal} {arXiv:2109.15209}\ } (\bibinfo {year}
  {2021})}\BibitemShut {NoStop}%
\bibitem [{\citenamefont {Werner}\ and\ \citenamefont
  {Eckstein}(2013)}]{werner_phonon-enhanced_2013}%
  \BibitemOpen
  \bibfield  {author} {\bibinfo {author} {\bibfnamefont {P.}~\bibnamefont
  {Werner}}\ and\ \bibinfo {author} {\bibfnamefont {M.}~\bibnamefont
  {Eckstein}},\ }\bibfield  {title} {\bibinfo {title} {Phonon-enhanced
  relaxation and excitation in the {Holstein}-{Hubbard} model},\ }\href
  {https://doi.org/10.1103/PhysRevB.88.165108} {\bibfield  {journal} {\bibinfo
  {journal} {Phys. Rev. B}\ }\textbf {\bibinfo {volume} {88}},\ \bibinfo
  {pages} {165108} (\bibinfo {year} {2013})}\BibitemShut {NoStop}%
\bibitem [{\citenamefont {Baroni}\ \emph {et~al.}(2001)\citenamefont {Baroni},
  \citenamefont {de~Gironcoli}, \citenamefont {Dal~Corso},\ and\ \citenamefont
  {Giannozzi}}]{baroni_phonons_2001}%
  \BibitemOpen
  \bibfield  {author} {\bibinfo {author} {\bibfnamefont {S.}~\bibnamefont
  {Baroni}}, \bibinfo {author} {\bibfnamefont {S.}~\bibnamefont
  {de~Gironcoli}}, \bibinfo {author} {\bibfnamefont {A.}~\bibnamefont
  {Dal~Corso}},\ and\ \bibinfo {author} {\bibfnamefont {P.}~\bibnamefont
  {Giannozzi}},\ }\bibfield  {title} {\bibinfo {title} {Phonons and related
  crystal properties from density-functional perturbation theory},\ }\href
  {https://doi.org/10.1103/RevModPhys.73.515} {\bibfield  {journal} {\bibinfo
  {journal} {Rev. Mod. Phys.}\ }\textbf {\bibinfo {volume} {73}},\ \bibinfo
  {pages} {515} (\bibinfo {year} {2001})}\BibitemShut {NoStop}%
\bibitem [{\citenamefont {Marini}\ \emph {et~al.}(2015)\citenamefont {Marini},
  \citenamefont {Ponc\'{e}},\ and\ \citenamefont
  {Gonze}}]{marini_many-body_2015}%
  \BibitemOpen
  \bibfield  {author} {\bibinfo {author} {\bibfnamefont {A.}~\bibnamefont
  {Marini}}, \bibinfo {author} {\bibfnamefont {S.}~\bibnamefont {Ponc\'{e}}},\
  and\ \bibinfo {author} {\bibfnamefont {X.}~\bibnamefont {Gonze}},\ }\bibfield
   {title} {\bibinfo {title} {Many-body perturbation theory approach to the
  electron-phonon interaction with density-functional theory as a starting
  point},\ }\href {https://doi.org/10.1103/PhysRevB.91.224310} {\bibfield
  {journal} {\bibinfo  {journal} {Phys. Rev. B}\ }\textbf {\bibinfo {volume}
  {91}},\ \bibinfo {pages} {224310} (\bibinfo {year} {2015})}\BibitemShut
  {NoStop}%
\bibitem [{\citenamefont {Lipavský}\ \emph {et~al.}(1986)\citenamefont
  {Lipavský}, \citenamefont {\v{S}pi\v{c}ka},\ and\ \citenamefont
  {Velický}}]{lipavsky_generalized_1986}%
  \BibitemOpen
  \bibfield  {author} {\bibinfo {author} {\bibfnamefont {P.}~\bibnamefont
  {Lipavský}}, \bibinfo {author} {\bibfnamefont {V.}~\bibnamefont
  {\v{S}pi\v{c}ka}},\ and\ \bibinfo {author} {\bibfnamefont {B.}~\bibnamefont
  {Velický}},\ }\bibfield  {title} {\bibinfo {title} {Generalized
  {Kadanoff}-{Baym} ansatz for deriving quantum transport equations},\ }\href
  {https://doi.org/10.1103/PhysRevB.34.6933} {\bibfield  {journal} {\bibinfo
  {journal} {Phys. Rev. B}\ }\textbf {\bibinfo {volume} {34}},\ \bibinfo
  {pages} {6933} (\bibinfo {year} {1986})}\BibitemShut {NoStop}%
\bibitem [{\citenamefont {Schl\"{u}nzen}\ \emph {et~al.}(2020)\citenamefont
  {Schl\"{u}nzen}, \citenamefont {Joost},\ and\ \citenamefont
  {Bonitz}}]{schlunzen_achieving_2020}%
  \BibitemOpen
  \bibfield  {author} {\bibinfo {author} {\bibfnamefont {N.}~\bibnamefont
  {Schl\"{u}nzen}}, \bibinfo {author} {\bibfnamefont {J.-P.}\ \bibnamefont
  {Joost}},\ and\ \bibinfo {author} {\bibfnamefont {M.}~\bibnamefont
  {Bonitz}},\ }\bibfield  {title} {\bibinfo {title} {Achieving the {Scaling}
  {Limit} for {Nonequilibrium} {Green} {Functions} {Simulations}},\ }\href
  {https://doi.org/10.1103/PhysRevLett.124.076601} {\bibfield  {journal}
  {\bibinfo  {journal} {Phys. Rev. Lett.}\ }\textbf {\bibinfo {volume} {124}},\
  \bibinfo {pages} {076601} (\bibinfo {year} {2020})}\BibitemShut {NoStop}%
\bibitem [{\citenamefont {Pavlyukh}\ \emph {et~al.}(2021)\citenamefont
  {Pavlyukh}, \citenamefont {Perfetto},\ and\ \citenamefont
  {Stefanucci}}]{pavlyukh_photoinduced_2021}%
  \BibitemOpen
  \bibfield  {author} {\bibinfo {author} {\bibfnamefont {Y.}~\bibnamefont
  {Pavlyukh}}, \bibinfo {author} {\bibfnamefont {E.}~\bibnamefont {Perfetto}},\
  and\ \bibinfo {author} {\bibfnamefont {G.}~\bibnamefont {Stefanucci}},\
  }\bibfield  {title} {\bibinfo {title} {Photoinduced dynamics of organic
  molecules using nonequilibrium {Green}'s functions with second-{Born},
  \textit{{GW}}, \textit{{T}}-matrix, and three-particle correlations},\ }\href
  {https://doi.org/10.1103/PhysRevB.104.035124} {\bibfield  {journal} {\bibinfo
   {journal} {Phys. Rev. B}\ }\textbf {\bibinfo {volume} {104}},\ \bibinfo
  {pages} {035124} (\bibinfo {year} {2021})}\BibitemShut {NoStop}%
\bibitem [{\citenamefont {Bittner}\ \emph {et~al.}(2021)\citenamefont
  {Bittner}, \citenamefont {Gole\v{z}}, \citenamefont {Casula},\ and\
  \citenamefont {Werner}}]{bittner_photoinduced_2021}%
  \BibitemOpen
  \bibfield  {author} {\bibinfo {author} {\bibfnamefont {N.}~\bibnamefont
  {Bittner}}, \bibinfo {author} {\bibfnamefont {D.}~\bibnamefont {Gole\v{z}}},
  \bibinfo {author} {\bibfnamefont {M.}~\bibnamefont {Casula}},\ and\ \bibinfo
  {author} {\bibfnamefont {P.}~\bibnamefont {Werner}},\ }\bibfield  {title}
  {\bibinfo {title} {Photoinduced {Dirac}-cone flattening in
  {BaNiS}$_{\textrm{2}}$},\ }\href
  {https://doi.org/10.1103/PhysRevB.104.115138} {\bibfield  {journal} {\bibinfo
   {journal} {Phys. Rev. B}\ }\textbf {\bibinfo {volume} {104}},\ \bibinfo
  {pages} {115138} (\bibinfo {year} {2021})}\BibitemShut {NoStop}%
\bibitem [{\citenamefont {Gole\v{z}}\ \emph {et~al.}(2019)\citenamefont
  {Gole\v{z}}, \citenamefont {Eckstein},\ and\ \citenamefont
  {Werner}}]{golez_multiband_2019}%
  \BibitemOpen
  \bibfield  {author} {\bibinfo {author} {\bibfnamefont {D.}~\bibnamefont
  {Gole\v{z}}}, \bibinfo {author} {\bibfnamefont {M.}~\bibnamefont
  {Eckstein}},\ and\ \bibinfo {author} {\bibfnamefont {P.}~\bibnamefont
  {Werner}},\ }\bibfield  {title} {\bibinfo {title} {Multiband nonequilibrium
  \textit{{GW}} + {EDMFT} formalism for correlated insulators},\ }\href
  {https://doi.org/10.1103/PhysRevB.100.235117} {\bibfield  {journal} {\bibinfo
   {journal} {Phys. Rev. B}\ }\textbf {\bibinfo {volume} {100}},\ \bibinfo
  {pages} {235117} (\bibinfo {year} {2019})}\BibitemShut {NoStop}%
\bibitem [{\citenamefont {Fehske}\ \emph {et~al.}(2011)\citenamefont {Fehske},
  \citenamefont {Wellein},\ and\ \citenamefont
  {Bishop}}]{fehske_spatiotemporal_2011}%
  \BibitemOpen
  \bibfield  {author} {\bibinfo {author} {\bibfnamefont {H.}~\bibnamefont
  {Fehske}}, \bibinfo {author} {\bibfnamefont {G.}~\bibnamefont {Wellein}},\
  and\ \bibinfo {author} {\bibfnamefont {A.~R.}\ \bibnamefont {Bishop}},\
  }\bibfield  {title} {\bibinfo {title} {Spatiotemporal evolution of polaronic
  states in finite quantum systems},\ }\href
  {https://doi.org/10.1103/PhysRevB.83.075104} {\bibfield  {journal} {\bibinfo
  {journal} {Phys. Rev. B}\ }\textbf {\bibinfo {volume} {83}},\ \bibinfo
  {pages} {075104} (\bibinfo {year} {2011})}\BibitemShut {NoStop}%
\bibitem [{\citenamefont {Sch\"{o}nhammer}(2019)}]{schonhammer_unusual_2019}%
  \BibitemOpen
  \bibfield  {author} {\bibinfo {author} {\bibfnamefont {K.}~\bibnamefont
  {Sch\"{o}nhammer}},\ }\bibfield  {title} {\bibinfo {title} {Unusual
  broadening of wave packets on lattices},\ }\href
  {https://doi.org/10.1119/1.5089752} {\bibfield  {journal} {\bibinfo
  {journal} {Am. J. Phys}\ }\textbf {\bibinfo {volume} {87}},\ \bibinfo {pages}
  {186} (\bibinfo {year} {2019})}\BibitemShut {NoStop}%
\bibitem [{\citenamefont {Claro}\ \emph {et~al.}(2003)\citenamefont {Claro},
  \citenamefont {Weisz},\ and\ \citenamefont
  {Curilef}}]{claro_interaction-induced_2003}%
  \BibitemOpen
  \bibfield  {author} {\bibinfo {author} {\bibfnamefont {F.}~\bibnamefont
  {Claro}}, \bibinfo {author} {\bibfnamefont {J.~F.}\ \bibnamefont {Weisz}},\
  and\ \bibinfo {author} {\bibfnamefont {S.}~\bibnamefont {Curilef}},\
  }\bibfield  {title} {\bibinfo {title} {Interaction-induced oscillations in
  correlated electron transport},\ }\href
  {https://doi.org/10.1103/PhysRevB.67.193101} {\bibfield  {journal} {\bibinfo
  {journal} {Phys. Rev. B}\ }\textbf {\bibinfo {volume} {67}},\ \bibinfo
  {pages} {193101} (\bibinfo {year} {2003})}\BibitemShut {NoStop}%
\bibitem [{\citenamefont {Balzer}\ \emph {et~al.}(2018)\citenamefont {Balzer},
  \citenamefont {Rasmussen}, \citenamefont {Schl\"{u}nzen}, \citenamefont
  {Joost},\ and\ \citenamefont {Bonitz}}]{balzer_doublon_2018}%
  \BibitemOpen
  \bibfield  {author} {\bibinfo {author} {\bibfnamefont {K.}~\bibnamefont
  {Balzer}}, \bibinfo {author} {\bibfnamefont {M.~R.}\ \bibnamefont
  {Rasmussen}}, \bibinfo {author} {\bibfnamefont {N.}~\bibnamefont
  {Schl\"{u}nzen}}, \bibinfo {author} {\bibfnamefont {J.-P.}\ \bibnamefont
  {Joost}},\ and\ \bibinfo {author} {\bibfnamefont {M.}~\bibnamefont
  {Bonitz}},\ }\bibfield  {title} {\bibinfo {title} {Doublon {Formation} by
  {Ions} {Impacting} a {Strongly} {Correlated} {Finite} {Lattice} {System}},\
  }\href {https://doi.org/10.1103/PhysRevLett.121.267602} {\bibfield  {journal}
  {\bibinfo  {journal} {Phys. Rev. Lett.}\ }\textbf {\bibinfo {volume} {121}},\
  \bibinfo {pages} {267602} (\bibinfo {year} {2018})}\BibitemShut {NoStop}%
\bibitem [{\citenamefont {Puig~von Friesen}\ \emph {et~al.}(2011)\citenamefont
  {Puig~von Friesen}, \citenamefont {Verdozzi},\ and\ \citenamefont
  {Almbladh}}]{puig_von_friesen_can_2011}%
  \BibitemOpen
  \bibfield  {author} {\bibinfo {author} {\bibfnamefont {M.}~\bibnamefont
  {Puig~von Friesen}}, \bibinfo {author} {\bibfnamefont {C.}~\bibnamefont
  {Verdozzi}},\ and\ \bibinfo {author} {\bibfnamefont {C.-O.}\ \bibnamefont
  {Almbladh}},\ }\bibfield  {title} {\bibinfo {title} {Can we always get the
  entanglement entropy from the {Kadanoff}-{Baym} equations? {The} case of the
  {T}-matrix approximation},\ }\href
  {https://doi.org/10.1209/0295-5075/95/27005} {\bibfield  {journal} {\bibinfo
  {journal} {Eurphys. Lett.}\ }\textbf {\bibinfo {volume} {95}},\ \bibinfo
  {pages} {27005} (\bibinfo {year} {2011})}\BibitemShut {NoStop}%
\bibitem [{\citenamefont {Stefanucci}\ and\ \citenamefont {van
  Leeuwen}(2013)}]{stefanucci_nonequilibrium_2013}%
  \BibitemOpen
  \bibfield  {author} {\bibinfo {author} {\bibfnamefont {G.}~\bibnamefont
  {Stefanucci}}\ and\ \bibinfo {author} {\bibfnamefont {R.}~\bibnamefont {van
  Leeuwen}},\ }\href {http://dx.doi.org/10.1017/CBO9781139023979} {\emph
  {\bibinfo {title} {Nonequilibrium {Many}-{Body} {Theory} of {Quantum}
  {Systems}: {A} {Modern} {Introduction}}}}\ (\bibinfo  {publisher} {Cambridge
  University Press},\ \bibinfo {address} {Cambridge},\ \bibinfo {year}
  {2013})\BibitemShut {NoStop}%
\bibitem [{\citenamefont {Sayyad}\ and\ \citenamefont
  {Eckstein}(2015)}]{sayyad_coexistence_2015}%
  \BibitemOpen
  \bibfield  {author} {\bibinfo {author} {\bibfnamefont {S.}~\bibnamefont
  {Sayyad}}\ and\ \bibinfo {author} {\bibfnamefont {M.}~\bibnamefont
  {Eckstein}},\ }\bibfield  {title} {\bibinfo {title} {Coexistence of excited
  polarons and metastable delocalized states in photoinduced metals},\ }\href
  {https://doi.org/10.1103/PhysRevB.91.104301} {\bibfield  {journal} {\bibinfo
  {journal} {Phys. Rev. B}\ }\textbf {\bibinfo {volume} {91}},\ \bibinfo
  {pages} {104301} (\bibinfo {year} {2015})}\BibitemShut {NoStop}%
\bibitem [{\citenamefont {Sch\"{u}ler}\ \emph {et~al.}(2018)\citenamefont
  {Sch\"{u}ler}, \citenamefont {Eckstein},\ and\ \citenamefont
  {Werner}}]{schuler_truncating_2018}%
  \BibitemOpen
  \bibfield  {author} {\bibinfo {author} {\bibfnamefont {M.}~\bibnamefont
  {Sch\"{u}ler}}, \bibinfo {author} {\bibfnamefont {M.}~\bibnamefont
  {Eckstein}},\ and\ \bibinfo {author} {\bibfnamefont {P.}~\bibnamefont
  {Werner}},\ }\bibfield  {title} {\bibinfo {title} {Truncating the memory time
  in nonequilibrium dynamical mean field theory calculations},\ }\href
  {https://doi.org/10.1103/PhysRevB.97.245129} {\bibfield  {journal} {\bibinfo
  {journal} {Phys. Rev. B}\ }\textbf {\bibinfo {volume} {97}},\ \bibinfo
  {pages} {245129} (\bibinfo {year} {2018})}\BibitemShut {NoStop}%
\bibitem [{\citenamefont {Kartsev}\ \emph {et~al.}(2014)\citenamefont
  {Kartsev}, \citenamefont {Verdozzi},\ and\ \citenamefont
  {Stefanucci}}]{kartsev_nonadiabatic_2014}%
  \BibitemOpen
  \bibfield  {author} {\bibinfo {author} {\bibfnamefont {A.}~\bibnamefont
  {Kartsev}}, \bibinfo {author} {\bibfnamefont {C.}~\bibnamefont {Verdozzi}},\
  and\ \bibinfo {author} {\bibfnamefont {G.}~\bibnamefont {Stefanucci}},\
  }\bibfield  {title} {\bibinfo {title} {Nonadiabatic {Van} der {Pol}
  oscillations in molecular transport},\ }\bibfield  {journal} {\bibinfo
  {journal} {Eur. Phys. J. B}\ }\textbf {\bibinfo {volume} {87}},\ \href
  {https://doi.org/10.1140/epjb/e2013-40905-5} {10.1140/epjb/e2013-40905-5}
  (\bibinfo {year} {2014})\BibitemShut {NoStop}%
\bibitem [{\citenamefont {Huber}\ \emph {et~al.}(2005)\citenamefont {Huber},
  \citenamefont {K\"{u}bler}, \citenamefont {T\"{u}bel}, \citenamefont
  {Leitenstorfer}, \citenamefont {Vu}, \citenamefont {Haug}, \citenamefont
  {K\"{o}hler},\ and\ \citenamefont {Amann}}]{huber_femtosecond_2005}%
  \BibitemOpen
  \bibfield  {author} {\bibinfo {author} {\bibfnamefont {R.}~\bibnamefont
  {Huber}}, \bibinfo {author} {\bibfnamefont {C.}~\bibnamefont {K\"{u}bler}},
  \bibinfo {author} {\bibfnamefont {S.}~\bibnamefont {T\"{u}bel}}, \bibinfo
  {author} {\bibfnamefont {A.}~\bibnamefont {Leitenstorfer}}, \bibinfo {author}
  {\bibfnamefont {Q.~T.}\ \bibnamefont {Vu}}, \bibinfo {author} {\bibfnamefont
  {H.}~\bibnamefont {Haug}}, \bibinfo {author} {\bibfnamefont {F.}~\bibnamefont
  {K\"{o}hler}},\ and\ \bibinfo {author} {\bibfnamefont {M.-C.}\ \bibnamefont
  {Amann}},\ }\bibfield  {title} {\bibinfo {title} {Femtosecond {Formation} of
  {Coupled} {Phonon}-{Plasmon} {Modes} in {InP}: {Ultrabroadband} {THz}
  {Experiment} and {Quantum} {Kinetic} {Theory}},\ }\href
  {https://doi.org/10.1103/PhysRevLett.94.027401} {\bibfield  {journal}
  {\bibinfo  {journal} {Phys. Rev. Lett.}\ }\textbf {\bibinfo {volume} {94}},\
  \bibinfo {pages} {027401} (\bibinfo {year} {2005})}\BibitemShut {NoStop}%
\bibitem [{\citenamefont {Dai}\ \emph {et~al.}(2015)\citenamefont {Dai},
  \citenamefont {Ma}, \citenamefont {Liu}, \citenamefont {Andersen},
  \citenamefont {Fei}, \citenamefont {Goldflam}, \citenamefont {Wagner},
  \citenamefont {Watanabe}, \citenamefont {Taniguchi}, \citenamefont
  {Thiemens}, \citenamefont {Keilmann}, \citenamefont {Janssen}, \citenamefont
  {Zhu}, \citenamefont {Jarillo-Herrero}, \citenamefont {Fogler},\ and\
  \citenamefont {Basov}}]{dai_graphene_2015}%
  \BibitemOpen
  \bibfield  {author} {\bibinfo {author} {\bibfnamefont {S.}~\bibnamefont
  {Dai}}, \bibinfo {author} {\bibfnamefont {Q.}~\bibnamefont {Ma}}, \bibinfo
  {author} {\bibfnamefont {M.~K.}\ \bibnamefont {Liu}}, \bibinfo {author}
  {\bibfnamefont {T.}~\bibnamefont {Andersen}}, \bibinfo {author}
  {\bibfnamefont {Z.}~\bibnamefont {Fei}}, \bibinfo {author} {\bibfnamefont
  {M.~D.}\ \bibnamefont {Goldflam}}, \bibinfo {author} {\bibfnamefont
  {M.}~\bibnamefont {Wagner}}, \bibinfo {author} {\bibfnamefont
  {K.}~\bibnamefont {Watanabe}}, \bibinfo {author} {\bibfnamefont
  {T.}~\bibnamefont {Taniguchi}}, \bibinfo {author} {\bibfnamefont
  {M.}~\bibnamefont {Thiemens}}, \bibinfo {author} {\bibfnamefont
  {F.}~\bibnamefont {Keilmann}}, \bibinfo {author} {\bibfnamefont {G.~C.
  A.~M.}\ \bibnamefont {Janssen}}, \bibinfo {author} {\bibfnamefont {S.-E.}\
  \bibnamefont {Zhu}}, \bibinfo {author} {\bibfnamefont {P.}~\bibnamefont
  {Jarillo-Herrero}}, \bibinfo {author} {\bibfnamefont {M.~M.}\ \bibnamefont
  {Fogler}},\ and\ \bibinfo {author} {\bibfnamefont {D.~N.}\ \bibnamefont
  {Basov}},\ }\bibfield  {title} {\bibinfo {title} {Graphene on hexagonal boron
  nitride as a tunable hyperbolic metamaterial},\ }\href
  {https://doi.org/10.1038/nnano.2015.131} {\bibfield  {journal} {\bibinfo
  {journal} {Nat. Nanotechnol.}\ }\textbf {\bibinfo {volume} {10}},\ \bibinfo
  {pages} {682} (\bibinfo {year} {2015})}\BibitemShut {NoStop}%
\bibitem [{\citenamefont {Verdi}\ \emph {et~al.}(2017)\citenamefont {Verdi},
  \citenamefont {Caruso},\ and\ \citenamefont {Giustino}}]{verdi_origin_2017}%
  \BibitemOpen
  \bibfield  {author} {\bibinfo {author} {\bibfnamefont {C.}~\bibnamefont
  {Verdi}}, \bibinfo {author} {\bibfnamefont {F.}~\bibnamefont {Caruso}},\ and\
  \bibinfo {author} {\bibfnamefont {F.}~\bibnamefont {Giustino}},\ }\bibfield
  {title} {\bibinfo {title} {Origin of the crossover from polarons to {Fermi}
  liquids in transition metal oxides},\ }\href
  {https://doi.org/10.1038/ncomms15769} {\bibfield  {journal} {\bibinfo
  {journal} {Nat. Commun.}\ }\textbf {\bibinfo {volume} {8}},\ \bibinfo {pages}
  {15769} (\bibinfo {year} {2017})}\BibitemShut {NoStop}%
\bibitem [{\citenamefont {Li}\ \emph {et~al.}(2021)\citenamefont {Li},
  \citenamefont {Trovatello}, \citenamefont {Dal~Conte}, \citenamefont {Nu\ss},
  \citenamefont {Soavi}, \citenamefont {Wang}, \citenamefont {Ferrari},
  \citenamefont {Cerullo},\ and\ \citenamefont
  {Brixner}}]{li_excitonphonon_2021}%
  \BibitemOpen
  \bibfield  {author} {\bibinfo {author} {\bibfnamefont {D.}~\bibnamefont
  {Li}}, \bibinfo {author} {\bibfnamefont {C.}~\bibnamefont {Trovatello}},
  \bibinfo {author} {\bibfnamefont {S.}~\bibnamefont {Dal~Conte}}, \bibinfo
  {author} {\bibfnamefont {M.}~\bibnamefont {Nu\ss}}, \bibinfo {author}
  {\bibfnamefont {G.}~\bibnamefont {Soavi}}, \bibinfo {author} {\bibfnamefont
  {G.}~\bibnamefont {Wang}}, \bibinfo {author} {\bibfnamefont {A.~C.}\
  \bibnamefont {Ferrari}}, \bibinfo {author} {\bibfnamefont {G.}~\bibnamefont
  {Cerullo}},\ and\ \bibinfo {author} {\bibfnamefont {T.}~\bibnamefont
  {Brixner}},\ }\bibfield  {title} {\bibinfo {title} {Exciton-phonon coupling
  strength in single-layer {MoSe}$_{\textrm{2}}$ at room temperature},\ }\href
  {https://doi.org/10.1038/s41467-021-20895-0} {\bibfield  {journal} {\bibinfo
  {journal} {Nat. Commun.}\ }\textbf {\bibinfo {volume} {12}},\ \bibinfo
  {pages} {954} (\bibinfo {year} {2021})}\BibitemShut {NoStop}%
\bibitem [{\citenamefont {Stefanucci}\ and\ \citenamefont
  {Perfetto}(2021)}]{stefanucci_carriers_2021}%
  \BibitemOpen
  \bibfield  {author} {\bibinfo {author} {\bibfnamefont {G.}~\bibnamefont
  {Stefanucci}}\ and\ \bibinfo {author} {\bibfnamefont {E.}~\bibnamefont
  {Perfetto}},\ }\bibfield  {title} {\bibinfo {title} {From carriers and
  virtual excitons to exciton populations: {Insights} into time-resolved
  {ARPES} spectra from an exactly solvable model},\ }\href
  {https://doi.org/10.1103/PhysRevB.103.245103} {\bibfield  {journal} {\bibinfo
   {journal} {Phys. Rev. B}\ }\textbf {\bibinfo {volume} {103}},\ \bibinfo
  {pages} {245103} (\bibinfo {year} {2021})}\BibitemShut {NoStop}%
\bibitem [{\citenamefont {Wu}\ \emph {et~al.}(2015)\citenamefont {Wu},
  \citenamefont {Basov},\ and\ \citenamefont {Fogler}}]{wu_topological_2015}%
  \BibitemOpen
  \bibfield  {author} {\bibinfo {author} {\bibfnamefont {J.-S.}\ \bibnamefont
  {Wu}}, \bibinfo {author} {\bibfnamefont {D.~N.}\ \bibnamefont {Basov}},\ and\
  \bibinfo {author} {\bibfnamefont {M.~M.}\ \bibnamefont {Fogler}},\ }\bibfield
   {title} {\bibinfo {title} {Topological insulators are tunable waveguides for
  hyperbolic polaritons},\ }\href {https://doi.org/10.1103/PhysRevB.92.205430}
  {\bibfield  {journal} {\bibinfo  {journal} {Phys. Rev. B}\ }\textbf {\bibinfo
  {volume} {92}},\ \bibinfo {pages} {205430} (\bibinfo {year}
  {2015})}\BibitemShut {NoStop}%
\bibitem [{\citenamefont {Murakami}\ \emph {et~al.}(2016)\citenamefont
  {Murakami}, \citenamefont {Werner}, \citenamefont {Tsuji},\ and\
  \citenamefont {Aoki}}]{murakami_multiple_2016}%
  \BibitemOpen
  \bibfield  {author} {\bibinfo {author} {\bibfnamefont {Y.}~\bibnamefont
  {Murakami}}, \bibinfo {author} {\bibfnamefont {P.}~\bibnamefont {Werner}},
  \bibinfo {author} {\bibfnamefont {N.}~\bibnamefont {Tsuji}},\ and\ \bibinfo
  {author} {\bibfnamefont {H.}~\bibnamefont {Aoki}},\ }\bibfield  {title}
  {\bibinfo {title} {Multiple amplitude modes in strongly coupled
  phonon-mediated superconductors},\ }\href
  {https://doi.org/10.1103/PhysRevB.93.094509} {\bibfield  {journal} {\bibinfo
  {journal} {Phys. Rev. B}\ }\textbf {\bibinfo {volume} {93}},\ \bibinfo
  {pages} {094509} (\bibinfo {year} {2016})}\BibitemShut {NoStop}%
\bibitem [{\citenamefont {Butler}\ \emph {et~al.}(2021)\citenamefont {Butler},
  \citenamefont {Yoshida}, \citenamefont {Hanaguri},\ and\ \citenamefont
  {Iwasa}}]{butler_doublonlike_2021}%
  \BibitemOpen
  \bibfield  {author} {\bibinfo {author} {\bibfnamefont {C.}~\bibnamefont
  {Butler}}, \bibinfo {author} {\bibfnamefont {M.}~\bibnamefont {Yoshida}},
  \bibinfo {author} {\bibfnamefont {T.}~\bibnamefont {Hanaguri}},\ and\
  \bibinfo {author} {\bibfnamefont {Y.}~\bibnamefont {Iwasa}},\ }\bibfield
  {title} {\bibinfo {title} {Doublonlike {Excitations} and {Their} {Phononic}
  {Coupling} in a {Mott} {Charge}-{Density}-{Wave} {System}},\ }\href
  {https://doi.org/10.1103/PhysRevX.11.011059} {\bibfield  {journal} {\bibinfo
  {journal} {Phys. Rev. X}\ }\textbf {\bibinfo {volume} {11}},\ \bibinfo
  {pages} {011059} (\bibinfo {year} {2021})}\BibitemShut {NoStop}%
\bibitem [{\citenamefont {Seibold}\ \emph {et~al.}(2014)\citenamefont
  {Seibold}, \citenamefont {B\"{u}nemann},\ and\ \citenamefont
  {Lorenzana}}]{seibold_time-dependent_2014}%
  \BibitemOpen
  \bibfield  {author} {\bibinfo {author} {\bibfnamefont {G.}~\bibnamefont
  {Seibold}}, \bibinfo {author} {\bibfnamefont {J.}~\bibnamefont
  {B\"{u}nemann}},\ and\ \bibinfo {author} {\bibfnamefont {J.}~\bibnamefont
  {Lorenzana}},\ }\bibfield  {title} {\bibinfo {title} {Time-{Dependent}
  {Gutzwiller} {Approximation}: {Interplay} with {Phonons}},\ }\href
  {https://doi.org/10.1007/s10948-013-2412-0} {\bibfield  {journal} {\bibinfo
  {journal} {J. Supercond. Nov. Magn.}\ }\textbf {\bibinfo {volume} {27}},\
  \bibinfo {pages} {929} (\bibinfo {year} {2014})}\BibitemShut {NoStop}%
\bibitem [{\citenamefont {Hirsch}\ and\ \citenamefont
  {Fradkin}(1983)}]{hirsch_phase_1983}%
  \BibitemOpen
  \bibfield  {author} {\bibinfo {author} {\bibfnamefont {J.~E.}\ \bibnamefont
  {Hirsch}}\ and\ \bibinfo {author} {\bibfnamefont {E.}~\bibnamefont
  {Fradkin}},\ }\bibfield  {title} {\bibinfo {title} {Phase diagram of
  one-dimensional electron-phonon systems. {II}. {The} molecular-crystal
  model},\ }\href {https://doi.org/10.1103/PhysRevB.27.4302} {\bibfield
  {journal} {\bibinfo  {journal} {Phys. Rev. B}\ }\textbf {\bibinfo {volume}
  {27}},\ \bibinfo {pages} {4302} (\bibinfo {year} {1983})}\BibitemShut
  {NoStop}%
\bibitem [{\citenamefont {De~Filippis}\ \emph {et~al.}(2012)\citenamefont
  {De~Filippis}, \citenamefont {Cataudella}, \citenamefont {Nowadnick},
  \citenamefont {Devereaux}, \citenamefont {Mishchenko},\ and\ \citenamefont
  {Nagaosa}}]{de_filippis_quantum_2012}%
  \BibitemOpen
  \bibfield  {author} {\bibinfo {author} {\bibfnamefont {G.}~\bibnamefont
  {De~Filippis}}, \bibinfo {author} {\bibfnamefont {V.}~\bibnamefont
  {Cataudella}}, \bibinfo {author} {\bibfnamefont {E.~A.}\ \bibnamefont
  {Nowadnick}}, \bibinfo {author} {\bibfnamefont {T.~P.}\ \bibnamefont
  {Devereaux}}, \bibinfo {author} {\bibfnamefont {A.~S.}\ \bibnamefont
  {Mishchenko}},\ and\ \bibinfo {author} {\bibfnamefont {N.}~\bibnamefont
  {Nagaosa}},\ }\bibfield  {title} {\bibinfo {title} {Quantum {Dynamics} of the
  {Hubbard}-{Holstein} {Model} in {Equilibrium} and {Nonequilibrium}:
  {Application} to {Pump}-{Probe} {Phenomena}},\ }\href
  {https://doi.org/10.1103/PhysRevLett.109.176402} {\bibfield  {journal}
  {\bibinfo  {journal} {Phys. Rev. Lett.}\ }\textbf {\bibinfo {volume} {109}},\
  \bibinfo {pages} {176402} (\bibinfo {year} {2012})}\BibitemShut {NoStop}%
\end{thebibliography}

\end{document}